\documentclass[]{aa}

\usepackage{graphicx}
\usepackage{natbib}
\usepackage{hyperref}
\hypersetup{pdfpagemode = {UseNone},
            pdftitle = {Migration jumps of planets in transition disks},
            pdfauthor = {Thomas Rometsch, Peter J. Rodenkirch, Wilhelm Kley, Cornelis P. Dullemond},
            pdfsubject = {},
            pdfview = {FitH},
            pdfstartview = {FitH},
            colorlinks = {true},
            linkcolor = [rgb]{0,0.35,0.7},
            citecolor = [rgb]{0,0.35,0.7},
            filecolor = [rgb]{0.61,0,0},
            urlcolor = [rgb]{0.61,0,0},
           }
\usepackage{txfonts}
\usepackage{enumitem}
\usepackage{placeins}

\providecommand{\defmodel}[1]{\textsf{#1}}
\providecommand{\model}[1]{\textsf{#1}}

\providecommand{\tabrefsB}{\ref{fig:migration-showcase}}
\providecommand{\tabrefsC}{\ref{fig:migration-resolution}}
\providecommand{\tabrefsD}{ \ref{fig:accretion-accreting}, \ref{fig:mass}}
\providecommand{\tabrefsE}{\ref{fig:migration-irradiation}, \ref{fig:hoverr-irradiation}}
\providecommand{\tabrefsF}{\ref{fig:migration-PDS70}}

\providecommand{\tabsep}{\\[-0.8em]\hline\\[-0.4em]}

\providecommand{\first}{1$^\text{st}$}
\providecommand{\second}{2$^\text{nd}$}
\providecommand{\third}{3$^\text{rd}$}
\providecommand{\fourth}{4$^\text{th}$}
\providecommand{\fifth}{5$^\text{th}$}

\begin{document}

\title{Migration jumps of planets in transition disks}

\author{Thomas Rometsch\inst{1}
  \and
  Peter J. Rodenkirch\inst{2}
  \and
  Wilhelm Kley\inst{1}
  \and
  Cornelis P. Dullemond\inst{2}
}

\institute{Institut für Astronomie und Astrophysik, Universit\"at T\"ubingen,
  Auf der Morgenstelle 10, 72076 T\"ubingen, Germany\\
  \email{thomas.rometsch@uni-tuebingen.de} \and Institute for Theoretical Astrophysics, Zentrum für Astronomie,
  Heidelberg University, Albert Ueberle Str. 2,
  69120 Heidelberg, Germany
}

\date{September 7, 2020}

\abstract
{
  Transition disks form a special class of protoplanetary disks that are characterized by a deficiency of
  disk material close to the star.
  In a subgroup, inner holes in these disks can stretch out to a few tens of au
  while there is still mass accretion onto the central star observed at the same time.
}
{
  We analyse the proposition that this type of wide transition disks is generated by the interaction
  of the disk with a system of embedded planets.
}
{
  We performed two-dimensional hydrodynamics simulations of a flat disk.
  Different equations of state were used including locally isothermal models and more realistic cases that consider viscous heating, radiative cooling and stellar heating.
  Two massive planets (with 3 to 9 Jupiter masses) were embedded in the disk and their dynamical evolution due to disk-planet
  interaction was followed for over 100\,000\,years.
  The simulations account for mass accretion onto the star and planets.
  We included models with parameters geared to the system PDS\,70.
  To assess the observability of features in our models we performed synthetic ALMA observations.
}
{
  For systems with a more massive inner planet,
  there are phases where both planets migrate outward engaged in a 2:1 mean motion resonance via the Masset-Snellgrove mechanism.
  In sufficiently massive disks the resulting formation of a vortex and the interaction with it can trigger rapid outward migration of the outer planet
  where its distance can increase by tens of au in a few thousand years. After another few thousand years, the outer planet
  rapidly migrates back inwards into resonance with the inner planet. We call this emerging composite phenomenon a {\it migration jump}.
  Outward migration and the migration jumps are accompanied by a high mass accretion rate onto the star.
  The synthetic images reveal numerous substructures depending on the type of dynamical behaviour. 
}
{
  Our results suggest that the outward migration of two embedded planets is a prime candidate for the explanation 
  of the observed high stellar mass accretion rate in wide transition disks.
  The models for PDS\,70 indicate it does not currently undergo a migration jump but might very well be in a phase of outward migration.
}

\keywords{accretion, accretion disks -- protoplanetary disks -- planet-disk interactions -- hydrodynamics -- methods: numerical}

\maketitle
%
%________________________________________________________________

\section{Introduction}

Observationally, transition disks are characterized by a lack of
flux in the few $\mu$-meter (near/mid IR) range as seen in the spectral energy distributions
(SEDs) of young stars. This flux deficit is typically associated with
'missing' dust having temperatures of 200-1000 K
\citep{2002ApJ...568.1008C,2005ApJ...621..461D}
corresponding to the inner regions of accretion disks. Despite this lack of dust,
there are nevertheless still signatures of gas accretion in several systems with large inner (dust) holes
that are a few tens of AU wide \citep[see e.g.][]{2014prpl.conf..497E}.

The observational properties of transitional disks (TDs) and previous modelling attempts
have been reviewed by \citet{2016PASA...33....5O} and we mention here only the main aspects relevant to this paper.
The origin of the inner disk clearing has been primarily attributed to three different processes:
photoevaporation from inside out through high energy radiation from the central young protostar
\citep[e.g.][]{1993Icar..106...92S,2006MNRAS.369..216A},
magnetically driven disk winds \citep[e.g.][]{2020A&A...633A..21R}
or by embedded massive companions
that carve deep gaps into the disk \citep[e.g.][]{2006ApJ...640.1110V}.
Additionally, TDs appear to come in two flavours,
mm-faint disks with low mm-fluxes, small inner holes ($\lesssim 10$au), and low accretion rates
onto the stars ($\approx 10^{-10} - 10^{-9}$ M$_\odot$/yr)
and mm-bright-disks with large mm-fluxes, large holes ($\gtrsim 20$au), and high accretion rates
$\approx 10^{-8}$ M$_\odot$/yr \citep{2012MNRAS.426L..96O} to which we refer here as
Type I and Type II disks, respectively.

While photoevaporation is certainly at work in some systems (Type I TDs), it is believed that it can only
operate for systems with a sufficiently low mass accretion rate below $10^{-8} M_\odot$/yr
and is otherwise quenched by the accretion flow \citep{2012MNRAS.426L..96O}.
At the same time the persistence of gas accretion within the inner (dust) holes is taken as an additional
indication that other mechanisms should operate that create these gaps \citep{2014A&A...568A..18M}.
The very likely mechanism for this second class of TDs is related to the growth of planets in the disks,
because young planets embedded in their nascent disks will not only
open a gap in the gas disk but they will create an even stronger
depletion of the dust near the planetary orbit \citep{2004A&A...425L...9P}.

Consequently, it has been suggested early on that the presence of a massive (Jupiter-sized) planet might be responsible for the
gap creation \citep{2006ApJ...640.1110V,2006MNRAS.373.1619R},
but at the same time it has been noticed that the gap created
by a single embedded planet is significantly narrower than observations of transition disks suggest.
Given the problems with a single planet and the photoevaporation models, it has been proposed that the main observational
features can be created by the presence of a system of (three to four) massive planets.
Following this line of thought, \citet{2011ApJ...729...47Z}
and \citet{2011ApJ...738..131D}
performed numerical simulations and argue that TDs are in fact {\itshape Signposts of young multi-planet systems}.
In this scenario the embedded planets act as a 'barrier' for the gas flow through the disk allowing some
gas to enter the inner region, causing the observed accretional features near the star, while the dust
is filtered out at the pressure maximum just beyond the outer edge of the gap and cannot enter the inner disk regions.
Following this line of thought, theoretical models with embedded planets and dust in disks
have been constructed to match the observed spectral energy distributions in the sub-millimeter
\citep{2013A&A...560A.111D,2015A&A...573A...9P}.

New ALMA observations focusing on CO-rotational lines have allowed
to determine the gas content in the inner disk region in more detail. These results show that
the inner disk gas depleted by factors of about $10^{2}$ \citep{2015A&A...579A.106V} or even to a factor
of $10^{4}$ with gas holes about a factor 2-3 smaller than the dust gaps \citep{2016A&A...585A..58V}
which is taken as another example of massive planets in disks \citep{2016Natur.530..169H}.
The conclusion that all or the majority of Type II TDs are shaped by massive planets has been
questioned by \citet{2016ApJ...825...77D} who argue that
there may not be enough giant planets to explain all observed Type II TDs, see also \citet{2008PASP..120..531C}
for the occurrence rate of massive planets at larger separations.
The solution to this problem is either that current numerical
models of planet-disk interactions are too inefficient at gap opening compared to Nature,
or that Type II TDs are intrinsically rare objects, rather than common and short-lived objects,
as is probably the case for their Type I counterparts. Considering that the arguments
of \citet{2016ApJ...825...77D} are based on analytical approximations of gap widths and sizes that
are based on isothermal disk models, it may well be that the theoretical models
have not reached the degree of sophistication necessary to produce reliable results.

A recent push to the planet based origin of Type II TDs came through the direct detection of embedded
planets in such systems.
For the T~Cha system the presence of a planet has been suggested based on observations in the L-band \citep{2011A&A...528L...7H}
which was later supported by ALMA observations, which indicate a gap in the disk ranging from 18 and 28\,au,
that are compatible with a $1.2 M_\mathrm{Jup}$ planet \citep{2018MNRAS.475L..62H}, but a direct confirmation is still pending.
A point like source has been detected in the L-band in the transition disk system MWC~758 
at a deprojected distance of about 20\,au from the star \citep{2018A&A...611A..74R}, 
directly hinting to the presence of an embedded planet.
The clearest evidence, however, comes from the PDS\,70 system. A first point like object was
detected in the near infrared at a projected distance of 22\,au, which was attributed
to a planet (PDS\,70b) orbiting within a gap that stretches from about 17 to 54\,au in size \citep{2018A&A...617A..44K}.
The large gap size in PDS\,70 hinted towards a second companion which was detected last year
\citep{2019NatAs...3..749H}. The authors confirmed the earlier H$\alpha$ detection of PDS\,70b and found a second
point-like H$\alpha$ source near the outer edge of the gap. This H$\alpha$ emission is taken as evidence for ongoing
accretion onto two proto-planets \citep{2019NatAs...3..749H}. In addition, the spatial separation of the two planets
indicate that they are close to a 2:1 mean motion resonance.

With respect to PDS\,70, a few simulations of embedded planets have been performed. The first study \citep{2019ApJ...879L...2M}
considered only one planet and a possible explanation of the wide gap was the creation of a large eccentric cavity
by a massive planet of about 2.5 $M_\mathrm{Jup}$. While in principle a possible scenario for sculpting transition disks,
as shown also by \citet{2013A&A...560A..40M},
the observation of a second planet rendered this scenario obsolete for PDS\,70. Consequently,
two-planet simulations were presented that show that a system engaged in a 2:1 mean motion resonance can in fact be stable
for several million of years \citep{2019ApJ...884L..41B}.

In this paper we study the evolution of planets embedded in protoplanetary disks using two-dimensional hydrodynamical simulations.
Our simulations include planet migration and mass accretion and either assume a locally isothermal equation of state
or incorporate stellar heating and radiative cooling from the disk.
The work extends an earlier study  \citep{2013A&A...560A..40M} where only one planet was considered that remained on a fixed
orbit around the star and was not allowed to migrate through the disk.
First, we will present generic models to demonstrate our new findings on the occurrence of migration jumps.
Then, we will study the specific system PDS\,70. For both cases we will generate synthetic images and discuss the observability of
the features.

In Sect.~\ref{sec:modelling}, we introduce our numerical model.
In Sect.~\ref{sec:results}, we present the evolution of the planetary system in our numerical simulations
and describe migration jumps in detail.
In Sect.~\ref{sec:observability}, we generate synthetic images of a disk in our simulations and identify possible observational features.
Sect.~\ref{sec:PDS70} is a case study of how migration jumps apply to the PDS\,70 system.
We discuss our findings in Sect.~\ref{sec:discussion} and give a summary in Sect.~\ref{sec:summary}.

\section{Modelling}
\label{sec:modelling}
In this section we describe the physical and numerical setup used in our simulations.
To give an overview of the cases investigated,
Table~\ref{tab:simulation_parameters} lists all models and summarizes their most important parameters.
Specific models are referred to by a \model{short label} which is typeset in sans serif characters.

\subsection{Physical setup}
We model an accretion disk around a young protostar solving the two-dimensional (2D, $r-\phi$)
viscous hydrodynamical equations, obtained by averaging over the vertical direction.
Most of our models assume a locally isothermal equation of state ($T = T(r)$ is constant over time).
For selected models (\model{IRR}, \model{PDS70 IRR}, \model{PDS70 IRR M/5})
we solve the energy equation and include heating by irradiation from the star (analogous to \cite{2020A&A...633A..29Z}),
viscous heating, and radiative cooling, using an averaged opacity.
All the details of the used set of equations are stated in \citet{2012A&A...539A..18M,2013A&A...560A..40M}.
Here, we do not solve for radiative transport within the plane of the disk.

For the systems investigated here, the locally isothermal assumption and the inclusion of the radiative effects yield
comparable results as the test in Appendix \ref{sec:depence_eos} shows.

Viscosity is parameterized with the $\alpha$ viscosity model \citep{1973A&A....24..337S} using a value $\alpha = 10^{-3}$.
The kinematic viscosity is then $\nu = \alpha c_s H$ with the sound speed $c_s$ and the disk's vertical pressure scale height $H$.
Together with the choices for $\Sigma$ and $H/r$ (as given below) this value corresponds to a viscous mass accretion
rate $\dot{M}_\text{disk} = 3\pi\Sigma\nu = 5.3\times 10^{-9}\,\text{M}_\odot/\text{yr}$ at 2\,au for the initial profile
(see also panel 2 of Fig.~\ref{fig:initial_conditions}).

The host star mass is $M_* = 1\,\text{M}_\odot$.
There are two planets embedded in the disk.
The inner planet (1) is initially located at $a_1 = 4\,a_\text{Jup} = 20.8\,\text{au}$
with a mass of $M_1 = 3$ to $9\,\text{M}_\text{Jup}$ and the outer planet (2) is initially located
at $a_2 = 7\,a_\text{Jup} = 36.4\,\text{au}$ with a mass of $M_2 = 1$ to $ 9\,\text{M}_\text{Jup}$.
These models are labelled with \model{M$k$-$l$} where $k$/$l$ is the mass of the inner/outer planet in $M_\text{Jup}$, respectively.
The exact combinations of masses can be found in Table \ref{tab:simulation_parameters}.
For all combinations of planet masses, a large deep gap can be expected in the disk.

To simplify the simulations, self-gravity of the disk is neglected in most of the simulations.
For the initial conditions of our standard model \model{M9-3} the Toomre $Q$ parameter is above 3 at 100\,au and 
above 8 close to the planet's location.
Thus, self-gravity should not play a dominant role in the system considered.
This assumption is verified by two additional models with self-gravity included.
The corresponding models are described in Sect.~\ref{sec:additional_models}.

\subsection{Simulation code}
\label{sec:simulation_code}
The hydrodynamics equations are solved on a 2D polar grid (using $r-\phi$ coordinates).
Radially, the domain ranges from $2.08$\,au to $208$\,au covered by 602 cells that are logarithmically spaced.
Azimuthally, the grid is uniformly spaced from $0$ to $2\pi$ with 821 cells.

We use a custom version of the FARGO code \citep{2000A&AS..141..165M}, which was also used by
\cite{2013A&A...560A..40M} in an earlier version,
and is further developed and maintained by our group at the University of T\"ubingen.
N-body calculations are performed with the IAS15 integrator in REBOUND \citep{2012A&A...537A.128R}
which is integrated into our version of FARGO.

\subsection{Initial conditions}

Initially, the disk's surface density $\Sigma$ is set to a power law profile of the form
\begin{align}\label{eqn:ic_sigma}
  \Sigma(r) = \Sigma_0 \left(\frac{r}{\text{au}}\right)^{-1} \,, \quad \Sigma_0 = 461.76\, \frac{\text{g}}{\text{cm}^2}\,.
\end{align}
The aspect ratio of the disk is chosen as $h=\frac{H}{r} = 0.05$ throughout the disk, which is equivalent to choosing a temperature power law of
\begin{align}\label{eqn:ic_T}
  T(r) = T_0 \left( \frac{r}{\mathrm{au}} \right)^{-1} \,, \quad T_0 = 632.86\, \text{K}\,.
\end{align}
We checked this assumption by performing additional simulations using irradiated disks, see
Appendix \ref{sec:depence_eos}.

\begin{figure}[t]
  \centering
  \includegraphics[width=\linewidth]{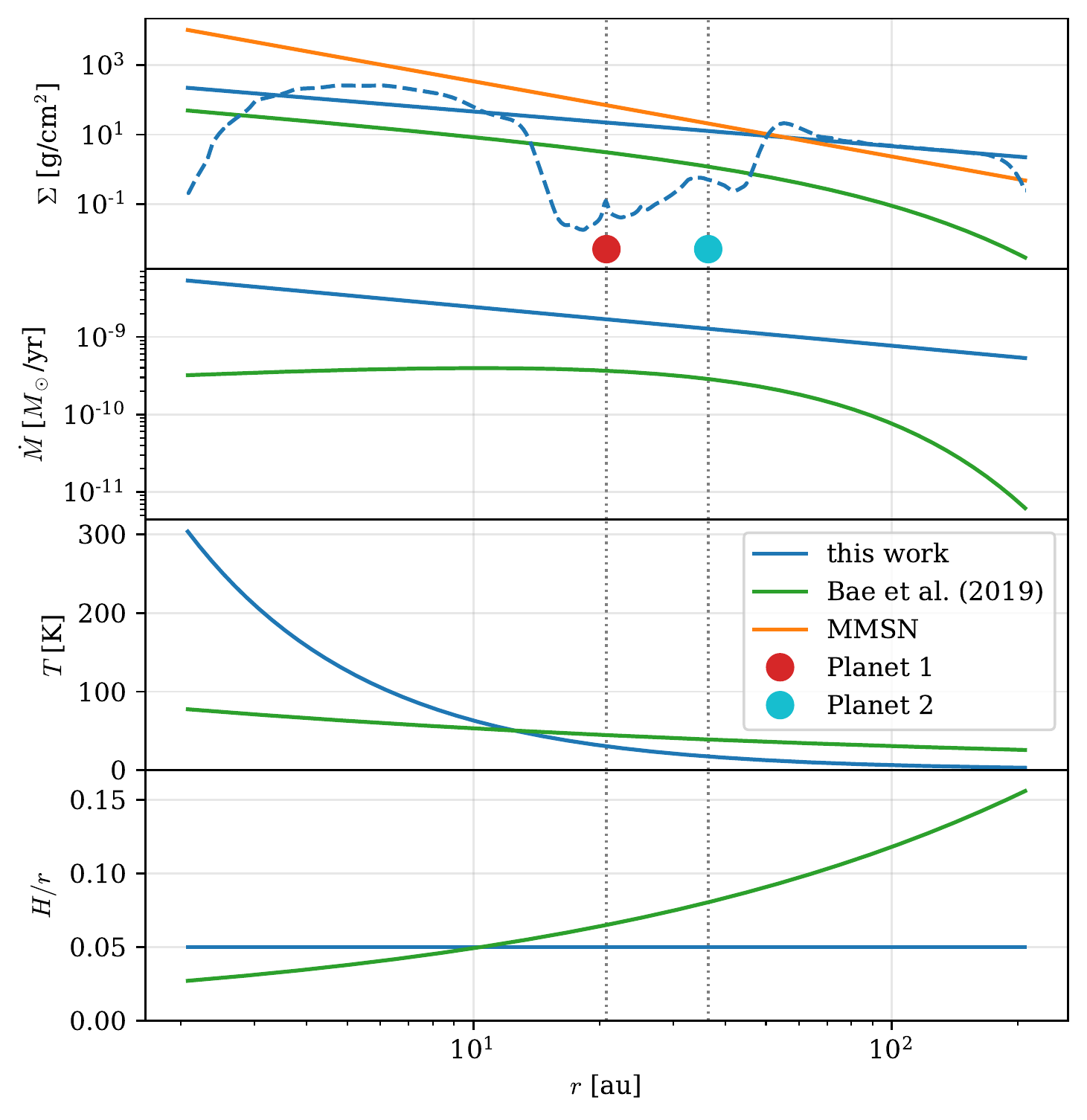}
  \caption{\label{fig:initial_conditions} Initial conditions for the disk around the $1\,\text{M}_\odot$ star
  used in our simulations (blue).
  These are compared to the initial conditions of the disk in \citet{2019ApJ...884L..41B}
  (green) in which the PDS70 system ($0.85\,\text{M}_\odot$ star) was modelled.
  For context,
  the surface density of the minimum mass solar nebula (MMSN) \citep{1981PThPS..70...35H} is plotted in orange.
  Physical initial conditions after the equilibration phase (see Fig.~\ref{fig:simulation_flow})
  at $t = 26\,\text{kyr}$ are shown for model \model{M9-3} as dashed blue line.
  The panels, from top to bottom, show the radial profile of the: surface density (top),
  viscous mass accretion rate, $\dot{M}_\text{disk} = 3\pi\Sigma\nu$ with $\alpha=10^{-3}$ (\second),
  temperature (\third) and aspect ratio (bottom).
  Initial location of the planets are marked by the vertical, dotted lines which span over all panels and
  are indicated by the red and cyan circles.}
\end{figure}

Both initial conditions are visualized in Fig.~\ref{fig:initial_conditions}.
The top panel shows $\Sigma(r)$.
Panel two shows the viscous mass accretion rate, $\dot{M}_\text{disk} = 3\pi\Sigma\nu$,
at each radius in the disk assuming $\alpha=10^{-3}$.
Panel three and the bottom panel show $T(r)$ and the resulting disk aspect ratio, $H/r$, respectively.
To establish a context, we also show the initial conditions of a PDS 70 simulation by \citet{2019ApJ...884L..41B} (B19) and
the minimum mass solar nebula (MMSN) \citep{1981PThPS..70...35H}.
Our disk has a lower $\Sigma$ than the MMSN in the inner $\approx 60\,\text{au}$
and is about 5 times larger than the $\Sigma$ in B19,
who model the 5.4\,Myr old PDS 70 system with a 0.85\,M$_\odot$ star.
Their disk is lighter, as it can be expected for an older system due to disk dispersal,
and has lower temperatures in the inner parts of the disk due to lower stellar luminosity.
The first panel also displays $\Sigma(r)$ after the initial equilibration phase as the dashed blue line.
Locations of the planets are marked by the dotted vertical lines and the red and cyan circles in the top panel.

\begin{figure}[t]
  \includegraphics[width=\linewidth]{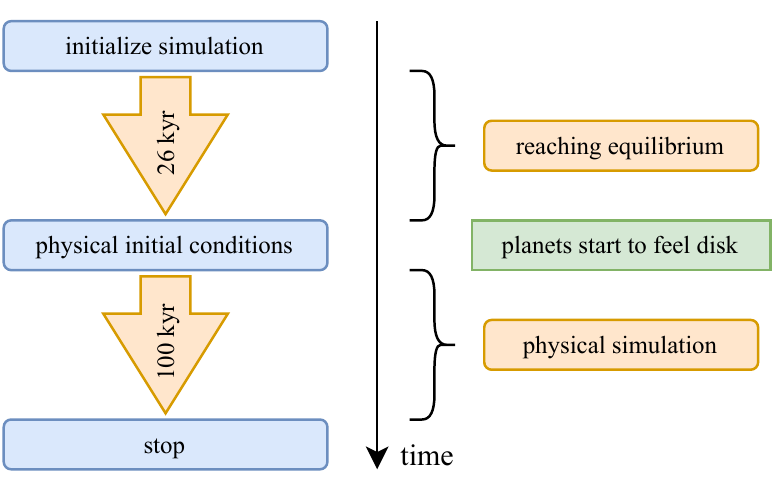}
  \caption{\label{fig:simulation_flow} Flow chart illustrating the equilibration and simulation phase.
    The disk properties are initialized according to the power laws in Eqs.~\eqref{eqn:ic_sigma} and \eqref{eqn:ic_T}.
    During the equilibration process (taking 26\,kyr, equivalent to
    274 orbits at the inner planet's initial location.) where the disk begins to 'feel' the planets,
    the density profile changed significantly,
    see the first panel of Fig.~\ref{fig:initial_conditions} for the difference, where the
    equilibrated profile is shown as the dashed blue line.}
\end{figure}

These power law initial conditions do not take into account the presence of embedded planets.
To obtain more physical initial conditions which are consistent with two massive embedded planets,
we insert the planets by ramping up their masses over $0.5\,\mathrm{kyr}$ while they are fixed at their orbits for
an initial equilibration time, $T_\text{eq}$.
During this time, the N-body system evolves without being subjected to the gravitational force of the disk while the
disk feels the N-body system and a common gap forms around the planets.
To determine $T_\text{eq}$, we analysed test simulations and checked for the time when the density waves caused by the
insertion of the planets left the computational domain
and the time after which the gap depth did not change significantly any more.
The second condition is fulfilled at a later time and yields $T_\text{eq} = 26\,\mathrm{kyr}$ which corresponds to
274 orbits at the inner planet's initial location.
After this time, the disk feedback onto star and planets is turned on which causes them to migrate.
The equilibration process is sketched by the flow chart in Fig. \ref{fig:simulation_flow}.

\subsection{Boundary conditions}
\label{sec:boundary_conditions}

We employ an outflow boundary (O) condition at both radial boundaries, $R_{min}$ and $R_{max}$.
This is done to let the disk evolve on its own, e.g. by allowing for an eccentric disk close to the inner boundary which
might be unphysically suppressed by the use of e.g. a wave damping boundary that imposes an azimuthal symmetry
close to the boundary.
Additionally, this choice of boundary condition allows density waves that are created by the insertion of the planets
to leave the computational domain.

For our outflow boundary condition, only mass flow leaving the domain is allowed.
It is implemented by enforcing a vanishing gradient of the energy density and $\Sigma$ and
by setting the radial velocity to zero at the boundary in the case that the velocity vector points into the domain
and by setting its gradient to zero otherwise.
This way, no mass can be generated.
The azimuthal velocity is left unchanged.

In order to check the validity of these boundary conditions for our physical setup, we studied other options.
This is important because the open boundary condition leads to an unphysical drop in surface density close to the boundary,
as it can be seen in the top panel of Fig.~\ref{fig:initial_conditions}.

One alternative is to employ an additional wave damping (WD) zone \citep{2006MNRAS.370..529D}
between $R_\text{min}$ and $2\,R_\text{min}$ where $\Sigma$ and the velocities are exponentially damped to their initial
values on a timescale of 5\% of the orbital timescale at $R_\text{min}$
(models \model{WD} at half resolution).
We combine this wave damping zone with the outflow boundary and with another common choice:
a closed, reflective boundary.
This reflective boundary does not allow mass flow through the boundary which is implemented by applying a zero gradient
for energy density and $\Sigma$ and by setting the ghost cells' radial velocities to zero at the boundary and to the negative of the
first active cell's radial velocity at the next ghost cell interface,
thus, essentially reflecting momentum at the boundary.
A combination of a reflective boundary and a wave damping zone is used in model \model{WDR}.

An outflow boundary as used in our simulations might overestimate the mass flow across the inner domain.
In a real system,
the properties of the accretion process onto the star are likely to limit the mass flow at the inner disk edge.
To test for the implication of this, we also employed a viscous internal boundary.
For this, the velocities of the inner ghost cells are set to a multiple of the viscous speed
\begin{align}\label{eqn:viscous_boundary}
  v(r_\text{in}) = \beta \, v_\text{visc}(r_\text{in})\,,\qquad v_\text{visc}(r_\text{in}) = -\frac{3\nu}{2r_\text{in}}
\end{align}
with the kinematic viscosity $\nu$.
We used $\beta = 1$ (model \model{VB}) and $\beta = 5$ (model \model{VB5}) which was found to yield a boundary comparable to simulations where the
2D grid is embedded in a larger 1D domain \citep{2007A&A...461.1173C}.
This boundary is also used to compare to similar models in the literature \citep[e.g.][]{2019AJ....157...45M}.

\subsection{Center of mass system}

Many simulations of planet-disk interactions use a grid centred on the primary star.
If there is only one gravitating object, this is an inertial frame.
With the addition of one or more planets, however,
it is not an inertial system any more and the so-called indirect term,
which is the negative of the force acting on the~star,
\begin{align}
  \vec{F}_\text{ind, star} = - \left( \vec{F}_\text{disk} + \vec{F}_\text{planets} \right) \,,
\end{align}
has to be applied to the bodies and the gas.
For more massive planets,
this causes the disk to oscillate for as much as the star oscillates around the center of mass,
which is undesirable from a numerical point of view.
For a full planetary system,
we obtain an inertial system by choosing the center of mass of the N-body system as the origin.
Then the indirect term vanishes except of the contribution from the disk which reads
\begin{align}
  \vec{a}_\text{ind., COM} = - \frac{1}{\sum_n M_n} \sum_n  M_n\,\vec{a}_\text{disk, n} \,,
\end{align}
for all N-body objects indexed by $n$ with mass $M_n$.
To avoid any numerical drift of the center of mass away from the origin,
we shift the whole N-body system every hydrodynamical timestep such that center of mass coincides with the origin,
see also \citet{2018A&A...616A..47T}.

\subsection{Gravitational interactions}

Star and planets are modelled as point masses.
The gravitational pull from the point masses onto the disk is implemented via their gravitational potential which is
\begin{align}
  \Psi(\vec{r}) = - \mathrm{G} \sum_n \frac{ M_n }{ \sqrt{ \left(\vec{r} - \vec{r}_n\right)^2 + \epsilon_\text{sm}^2  } }
\end{align}
at location $\vec{r}$, and the index $n$ runs over all point masses with masses $M_n$ and position vectors $\vec{r}_n$.
The smoothing length is chosen to be $\epsilon_\text{sm} = 0.6 H(\vec{r})$, with the local disk scale height $H(\vec{r})$, to approximate the gravitational force in a 3D disk \citep{2012A&A...541A.123M}.

The back reaction from the disk onto the point masses is calculated by direct summation over all grid cells,
which are indexed by $k$, have masses $m_k$ and positions $\vec{r_k}$.
The total disk force acting on a point mass with index $n$ is
\begin{align}
  \vec{F}_\text{disk, n} = - \mathrm{G} M_n \sum_k \frac{m_k}{\left( \vec{r_n} - \vec{r_k} \right)^2 + \epsilon_\text{sm}^2} \frac{\vec{r_n} - \vec{r_k}}{ \left| \vec{r_n} - \vec{r_k} \right| } \,,
\end{align}
where $\epsilon_\text{sm}$ is the same as used for the potential.

\subsection{Implementation of planetary accretion}
In some simulations we include the possibility of mass accretion onto the embedded planets.
We use the prescription from \cite{2017A&A...598A..80D} and remove a fraction of mass every time step $\Delta t$
from the hydro simulation in the vicinity of the planet.
Close to the planet, more mass is removed than further away.
No mass is removed beyond a distance of $0.5 R_\text{Hill}$ from the planet's location. For full
details see \citet{2017A&A...598A..80D}.
The mass removed at each time step follows the relation
\begin{align}\label{eqn:mass_accretion_planet}
  \Delta M & = f_\text{acc}\, M_\text{vicinity} \, \Delta t \,\Omega_\text{K} \,,
\end{align}
where $f_\text{acc}$ is a free parameter to control the efficiency of the accretion process.
We use values of $f_\text{acc} = 10^{-4}, 10^{-3}, 10^{-2}$ and $10^{-1}$
(models \model{A$f_\text{acc}$} with $f_\text{acc}$ in decimal notation).
Mass and angular momentum is conserved by adding the mass removed from the hydro simulation to the planet's mass
and adding an equivalent amount of angular momentum.

\subsection{Additional models}
\label{sec:additional_models}

In order to test our model choices, we ran additional simulations with different parameters.

To check the impact of a constant aspect ratio, we rerun the standard model \model{M9-3} with a flaring aspect ratio of
$h(r) = 0.019\,\left( \frac{r}{\mathrm{au}} \right)^{2/7}$ (model \model{FLARE}).
This flaring corresponds to a disk dominated by irradiation and $h=0.05$ is reached at $30\,\mathrm{au}$.

A resolution test was carried out by lowering/increasing the resolution by a factor of 2 ($\sqrt{2}$ in each direction,
models \model{M9-3 HR} and \model{M9-3 DR}).
See appendix~\ref{sec:resolution_test} for the comparison.

To test the influence of the domain size, half resolution models (same $\Delta r/r$ as \model{M9-3 HR}) were done with
a larger grid spanning from $5.2\,\text{au}$ to $520.0\,\text{au}$ (models \model{L}, \model{L M/2}, \model{L M/10}).
For the larger domain, density waves which are caused by the insertion of planets take longer to leave the domain.
Therefore, $T_\text{eq} = 59\,\text{kyr}$ is chosen.

Some models were run with a lower surface density but otherwise identical parameters to their sibling models.
They are indicated by the \model{M/$N$} in their label and are initialized with $N$ times smaller surface density
compared to Eq.~\eqref{eqn:ic_T}
(models \model{M9-3 M/10}, \model{L M/2}, \model{L M/10} and \model{PDS70 IRR M/5}).

To test the effect of the center of mass frame,
we repeated some models with a coordinate system centered on the primary star.
These are indicated by a \model{-P} in their labels (models \model{VB-P}, \model{VB5-P}, \model{WD-P}, \model{WDR-P}).

Most of our models do not include self-gravity.
To test the implications of self-gravity, we ran two additional models with self-gravity included.
It is implemented analogously to \cite{baruteau...thesis}, but with a 
modified smoothing length that includes a dependence on radius 
in order to fulfil Newton's third law (Moldenhauer \& Kley, in preparation).
The two models are based on model \model{VB5-P} for runtime reasons because this model has the largest timestep.
Model \model{SG} has just self-gravity enabled
 and \model{SG IRR} additionally solves the energy equation and considers irradiation like model \model{IRR}.

We also ran a set of models to simulate the PDS 70 system.
The parameters and results are discussed separately in Sect.~\ref{sec:PDS70}.

\begin{table*}[htb]
  \begin{center}
    \caption{\label{tab:simulation_parameters}Model parameters and outcome of the simulations.}
    \renewcommand\arraystretch{0.8}
    % \small
    \begin{tabular}{lccccclcccccc}
      \hline\hline                                                                                                                                                     \\[-0.4em]
      Label
                               & Bound\,\tablefootmark{a}
                               & Res\,\tablefootmark{b}
                               & $M_1$\,\tablefootmark{c}
                               & $M_2$\,\tablefootmark{c}
                               & $f_\text{acc}$
                               & $\Sigma$\,\tablefootmark{d}
                               &
                               & Migration\,\tablefootmark{e}
                               & Jump\,\tablefootmark{f}
                               & Events\,\tablefootmark{g}
                               & Fig.                                                                                                                                  \\[0.2em]
      \hline                                                                                                                                                           \\[-0.4em]
      \defmodel{M9-3, A0.0}    & O                            &     & 9 & 3   &           &      &  & $\nearrow$            & $\checkmark$ &    & \tablefootmark{h}            \\
      \defmodel{M6-2}          & O                            &     & 6 & 2   &           &      &  & $\nearrow$            & $\checkmark$ &    &                      \\
      \defmodel{M3-1}          & O                            &     & 3 & 1   &           &      &  & $\nearrow$            & $\checkmark$ &    &                      \\
      \tabsep
      \defmodel{M9-4.5}        & O                            &     & 9 & 4.5 &           &      &  & $\nearrow$            & $\checkmark$ & DS & \tabrefsB            \\
      \defmodel{M6-3}          & O                            &     & 6 & 3   &           &      &  & $\nearrow$            & $\checkmark$ &    &                      \\
      \defmodel{M6-6}          & O                            &     & 6 & 6   &           &      &  & $\searrow$            &              &    & \tabrefsB            \\
      \tabsep
      \defmodel{M3-9}          & O                            &     & 3 & 9   &           &      &  & $\searrow$            &              &    & \tabrefsB            \\
      \defmodel{M2-6}          & O                            &     & 2 & 6   &           &      &  & $\searrow$            &              &    &                      \\
      \defmodel{M3-1.5}        & O                            &     & 3 & 1.5 &           &      &  & $\searrow$            &              &    &                      \\
      \tabsep
      \defmodel{M9-3 HR}       & O                            & 1/2 & 9 & 3   &           &      &  & $\nearrow$            & $\checkmark$ &    & \tabrefsC            \\
      \defmodel{M9-3 DR}       & O                            & 2   & 9 & 3   &           &      &  & $\nearrow$            & $\checkmark$ &    & \tabrefsC            \\
      \defmodel{M9-3 M/10}     & O                            &     & 9 & 3   &           & 1/10 &  & $\nearrow$            &              &    &                      \\
      \tabsep
      \defmodel{IRR}           & O                            &     & 9 & 3   &           &      &  & $\nearrow$            & $\checkmark$ &    & \tabrefsE            \\
      \defmodel{FLARE}         & O                            &     & 9 & 3   &           &      &  & $\nearrow$            & $\checkmark$ &    & \tabrefsE            \\
      \tabsep
      \defmodel{A0.0001}       & O                            &     & 9 & 3   & $10^{-4}$ &      &  & $\nearrow$            & $\checkmark$ &    & \tabrefsD            \\
      \defmodel{A0.001}        & O                            &     & 9 & 3   & $10^{-3}$ &      &  & $\nearrow$            & $\checkmark$ &    & \tabrefsD            \\
      \defmodel{A0.01}         & O                            &     & 9 & 3   & $10^{-2}$ &      &  & $\nearrow$            & $\checkmark$ & DS & \tabrefsD            \\
      \defmodel{A0.1}          & O                            &     & 9 & 3   & $10^{-1}$ &      &  & $\nearrow$            & $\checkmark$ &    & \tabrefsD            \\
      \tabsep
      \defmodel{L}             & O                            & 1/2 & 9 & 3   &           &      &  & $\nearrow$ $\searrow$ & $\checkmark$ & S  & \ref{fig:orbit-swap} \\
      \defmodel{L M/2}         & O                            & 1/2 & 9 & 3   &           & 1/2  &  & $\nearrow$            & $\checkmark$ &    &                      \\
      \defmodel{L M/10}        & O                            & 1/2 & 9 & 3   &           & 1/10 &  & $\nearrow$            &              &    &                      \\
      \tabsep
      \defmodel{VB}            & V                            &     & 9 & 3   &           &      &  & $\nearrow$            & $\checkmark$ & E  &                      \\
      \defmodel{VB5}           & V5                           &     & 9 & 3   &           &      &  & $\nearrow$            & $\checkmark$ & E  &                      \\
      \defmodel{WD}            & O+WD                         & 1/2 & 9 & 3   &           &      &  & $\nearrow$            & $\checkmark$ & E  &                      \\
      \defmodel{WDR}           & R+WD                         & 1/2 & 9 & 3   &           &      &  & $\nearrow$            & $\checkmark$ & E  &                      \\
      \tabsep
      \defmodel{VB-P}          & V                            &     & 9 & 3   &           &      &  & $\nearrow$            & $\checkmark$ &    &                      \\
      \defmodel{VB5-P}         & V5                           &     & 9 & 3   &           &      &  & $\nearrow$            & $\checkmark$ &    &                      \\
      \defmodel{WD-P}          & O+WD                         & 1/2 & 9 & 3   &           &      &  & $\nearrow$            & $\checkmark$ &    &                      \\
      \defmodel{WDR-P}         & R+WD                         & 1/2 & 9 & 3   &           &      &  & $\nearrow$            & $\checkmark$ &    &                      \\
      \tabsep
      \defmodel{SG}           & V5                            &     & 9 & 3   &           &      &  & $\nearrow$ $\searrow$ & $\checkmark$ & S   &                      \\
      \defmodel{SG IRR}       & V5                            &     & 9 & 3   &           &      &  & $\nearrow$ $\searrow$ & $\checkmark$ & DS, S   &                      \\
      \tabsep
      \defmodel{PDS70 ISO}     & O                            &     & 9 & 3   &           &      &  & $\nearrow$            & $\checkmark$ &    & \tabrefsF            \\
      \defmodel{PDS70 IRR}     & O                            &     & 9 & 3   &           &      &  & $\nearrow$            & $\checkmark$ &    & \tabrefsF            \\
      \defmodel{PDS70 IRR M/5} & O                            &     & 9 & 3   &           & 1/5  &  & $\nearrow$            &              &    & \tabrefsF            \\[0.2em]
      \hline
    \end{tabular}
  \end{center}
  \tablefoot{
    If a field is empty, the values are the same as for the reference model \model{M9-3}.
    \tablefoottext{a}{For the inner boundary, the following choices are possible: O means an outflow boundary,
      R a closed, reflective boundary, V a viscous boundary, V5 a viscous boundary with 5 times viscous speed,
      and +WD indicates an additional wave damping zone close to the inner boundary,
      see Sect.~\ref{sec:boundary_conditions} for more detail.}
    \tablefoottext{b}{The factor refers to the 2D resolution compared to the \model{M9-3} case
      (Sect.~\ref{sec:simulation_code}).}
    \tablefoottext{c}{Planet masses in units of $M_\text{Jup}$.}
    \tablefoottext{d}{Surface density in units of the reference density $\Sigma_0$ (see Eq.~\eqref{eqn:ic_sigma}).}
    \tablefoottext{e}{Direction of migration where $\nearrow$ and $\searrow$ indicate outward and inward migration, respectively.}
    \tablefoottext{f}{$\checkmark$, if at least one migration jump happened in the simulation.}
    \tablefoottext{g}{Additional events of interest where S indicates a single orbit swap,
      DS a double orbit swap and E stands for planet ejection.}
    \tablefoottext{h}{The standard model \model{M9-3} is used in Fig.\ \ref{fig:migration-showcase}, 
      \ref{fig:migration-sample-zoomin}, \ref{fig:2d_surface_density}, \ref{fig:mass}, \ref{fig:migration-resolution}, 
      \ref{fig:migration-irradiation} \ref{fig:hoverr-irradiation}, \ref{fig:streamlines} 
      and \ref{fig:vortensity}.}
    }
\end{table*}

\section{Results}
\label{sec:results}

In this section, we analyse the simulations listed in Table~\ref{tab:simulation_parameters}
with respect to their dynamical evolution and accretion properties.
Each effect is described in a separate subsection.

Due to the large number of simulations performed, we do not visualize all of them.
Instead, we used a representative selection to highlight the various dynamical evolutions.
If not stated otherwise, the features observed in those simulations that are not displayed are very similar but might happen
at a different point in time.

The simulation outcomes are described in the following and an overview can be found
in Table~\ref{tab:simulation_parameters}.
The different properties listed are: outward and inward migration ($\nearrow$ and $\searrow$, Sect. \ref{sec:direction_migration}),
migration jumps (indicated by $\checkmark$, Sect.~\ref{sec:migration_jumps}), single/double orbit swaps (S/DS, Sect.~\ref{sec:orbit_swaps}),
and planet ejections (E, Sect.~\ref{sec:planet_ejection}).

Table~\ref{tab:simulation_parameters} also refers to the corresponding figures which show the migration history of the
respective simulations.

\subsection{Direction of migration}
\label{sec:direction_migration}

\begin{figure}[t]
  \centering
  \includegraphics[width=\linewidth]{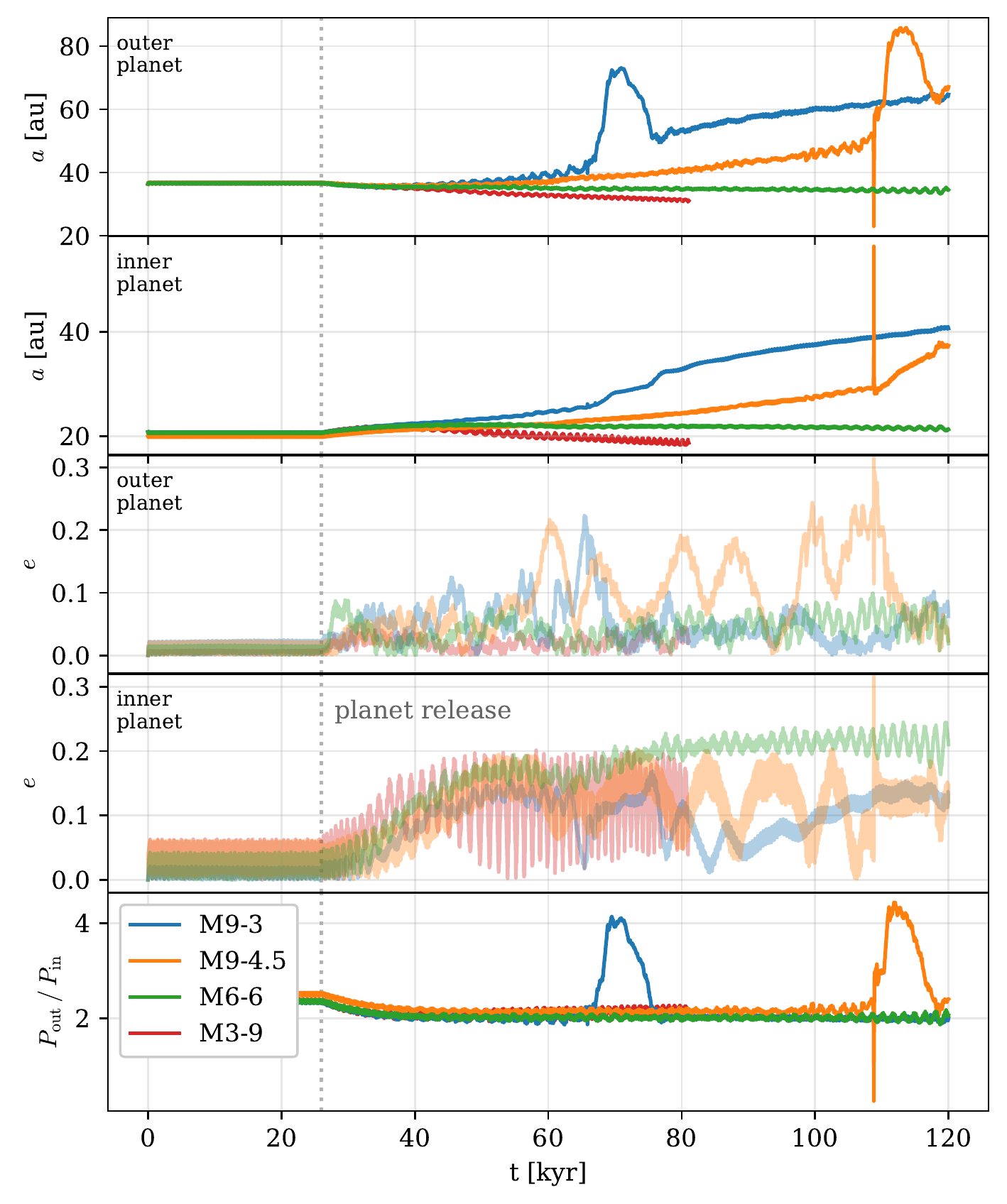}
  \caption{\label{fig:migration-showcase}Migration history for a selection of models described in
    Table~\ref{tab:simulation_parameters} to highlight the different possible dynamical evolution of the embedded planets.
    The panels show, from top to bottom, the evolution of: a) the semi-major axis, $a$, of outer/inner planet,
    b) their eccentricities, $e$, and c) their period ratio.
    The release time of the planets is marked by the vertical dotted line.
    Most prominent is the occurrence of fast outward migration ($migration jump$) for model
    \model{M9-3} (blue) between 60\,kyr and 80\,kyr, and a double orbit swap at 110\,kyr
    followed by a migration jump for model \model{M9-4.5}.
    At the time of the double orbit swap at $t \sim 110$\,kyr, $e_\text{in} =0.38$ and $e_\text{out}= 0.58$.
    They are cut out to increase visibility of the rest of the data.
  }
\end{figure}

In all simulations,
the planets start to migrate after gravitational feedback from the disk onto the planets is turned on.
A selection of migration tracks for simulations with different planetary masses and mass ratios is shown in
Fig.~\ref{fig:migration-showcase}.
The selection showcases the different possible behaviours of the systems.

For all combinations of planetary masses, the inner planet is migrating outward for the first $10\,\mathrm{kyr}$ because it only feels the positive torque contribution from the inner disk.
There is no substantial contribution to the torque from the outer disk due to the large common gap open by the two planets.
Regions where the disk's interaction would be strongest \citep{1997Icar..126..261W} are cleared by the outer planet
\citep{2007A&A...472..981S}.
Vice versa, the outer planet migrates inward for the first $10\,\mathrm{kyr}$.
After this time, the planets are captured in a 2:1 mean motion resonance (MMR).
We verified this by checking the resonant angles for an inner mean motion resonance, $\Theta_{1,2}^{2:1} = 2\lambda_2 - \lambda_1 - \varpi_{1,2}$
\citep{2018MNRAS.477.3383F}, where $\lambda = M + \varpi$ is the mean orbital longitude with the mean anomaly, $M$, and
$\varpi = \omega + \Omega$ is the longitude of periapsis with the argument of periapsis, $\omega$, and the longitude
of the ascending node, $\Omega$.
These angles are indeed librating around zero (see Fig. \ref{fig:migration-sample-zoomin} where
$\Theta_{1,2}^{2:1}$ are displayed).
Being locked in resonance, the planets then migrate in unison with a direction depending on
whether the positive torque contribution of the inner disk (inside the common gap) is larger in magnitude 
than the negative torque contribution from the outer disk (outside the common gap).
Since these torque contributions depend on the respective planetary mass, the direction of migration can vary.
If the inner planet is more massive than the outer planet, the system can migrate outwards as found initially for the Jupiter-Saturn
system by \citet{2001MNRAS.320L..55M}.
Indeed,
Fig.~\ref{fig:migration-showcase} shows outward migration for the \model{M9-3} and \model{M9-4.5} models
for which the inner planet is more massive than the outer one.
Conversely, the planets migrate inward for a more massive outer planet as in model \model{M3-9}.
For equal mass planets (\model{M6-6}), the system still migrates inward, but with a lower migration rate.

\subsection{Migration Jumps\label{sec:migration_jumps}}

Simulations in which the inner planet is more massive and which show outward migration occasionally
exhibit an additional process which we call a 'migration jump':
 
\begin{enumerate}
  \item The outer planet embarks on a rapid outward migration covering tens of au in a time period of a few kyr.
  \item After reaching a maximum radius, it stays in this region for several kyr, occasionally up to tens of kyr.
  \item It migrates back inward, again on a short timescale,
        until it locks back into resonance with the inner planet.
\end{enumerate}
Two examples of a migration jump can be seen in Fig.~\ref{fig:migration-showcase}.
This process can repeat itself multiple times (see Fig.~\ref{fig:migration-irradiation}).

\begin{figure}[t]
  \centering
  \includegraphics[width=\linewidth]{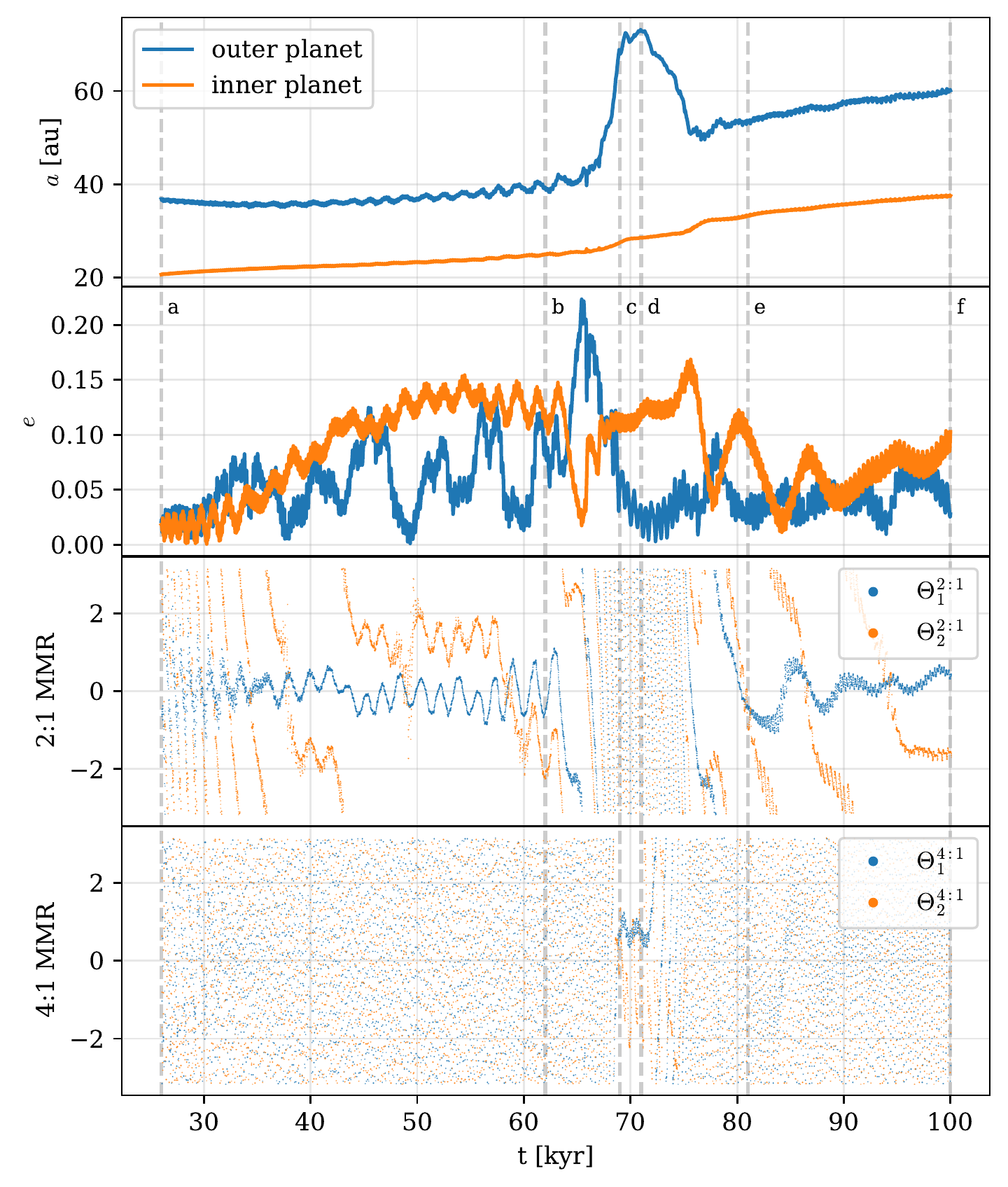}
  \caption{\label{fig:migration-sample-zoomin}Zoom-in on a migration jump in model \model{M9-3}
    including the time leading up to and following the event.
    The panels show, from top to bottom, the evolution of: semi-major axis of both planets (top),
    their eccentricities (\second), 2:1 MMR resonant angles (\third),
    and 4:1 MMR resonant angles (bottom).
    The vertical lines correspond to the snapshots shown in Fig. \ref{fig:2d_surface_density}.}
\end{figure}

\begin{figure*}[t]
  \centering
  \includegraphics[width=\linewidth]{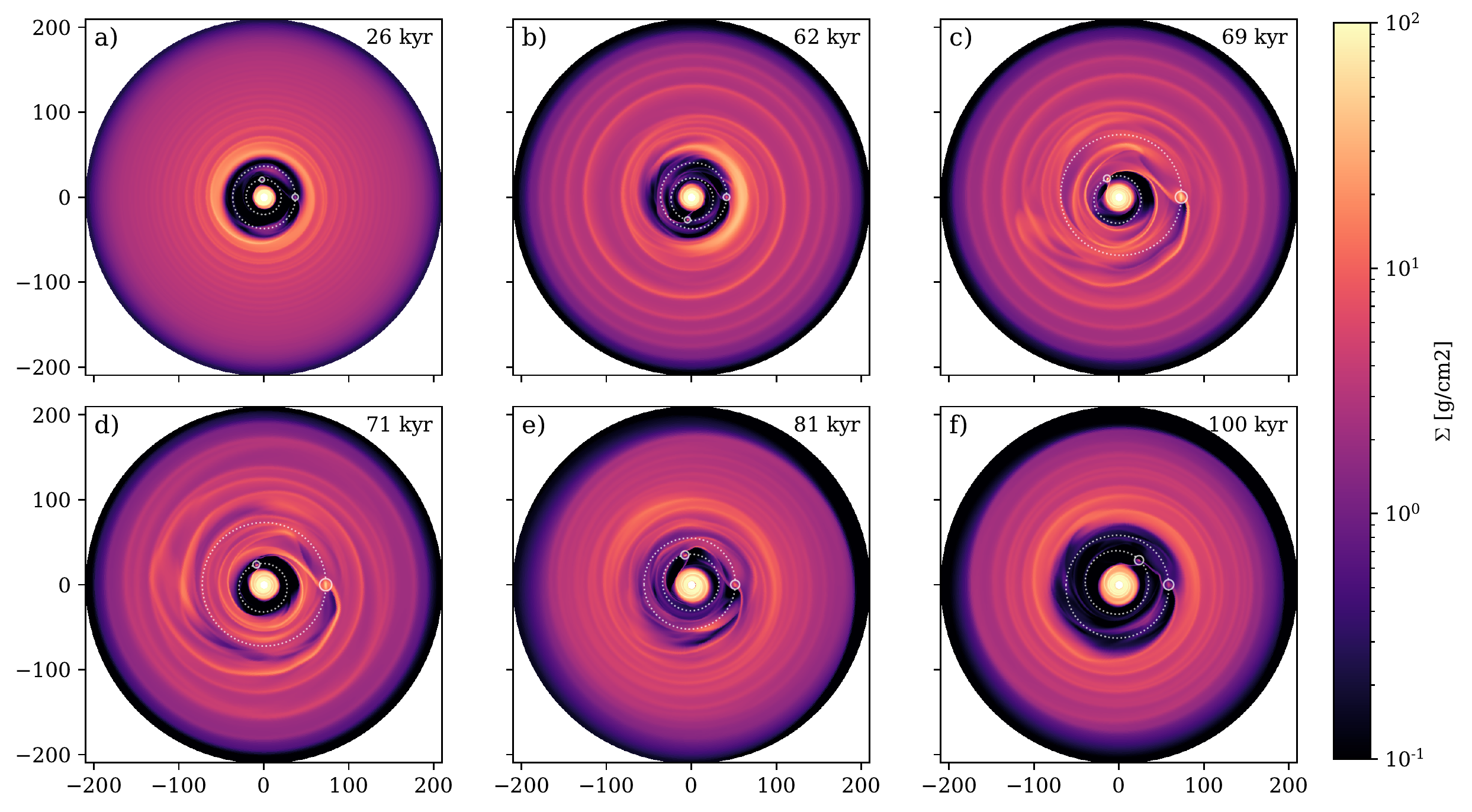}
  \caption{\label{fig:2d_surface_density}Snapshots of the surface density for model \model{M9-3}
    showing the disk at different times: prior (a and b), during (c and d), and after (e and f) a migration jump.
    The current orbits of the two planets are marked by the dotted white ellipses and the planetary Hill spheres are
    indicated by the small circles.
    Coordinate labels show the position in au.
    The snapshots are rotated to have the outer planet fixed on the horizontal axis to the right of the origin.
    Time inside the simulation is shown in the upper right corner.
    To locate the time of a snapshot in the timeline of the simulation, refer to the annotated vertical lines in
    Fig.~\ref{fig:migration-sample-zoomin}.
    For synthetic observations of the snapshots, refer to Fig.~\ref{fig:synth_obs_montage}.}
\end{figure*}

Figure~\ref{fig:migration-sample-zoomin} shows a zoom on one migration jump of our standard \model{M9-3} model focussed
on the time leading up to the event and some time afterwards.
It shows from top to bottom: the semi-major axis of both planets (top), their eccentricities (\second),
the 2:1 MMR angles $\Theta_{1,2}^{2:1} = 2\lambda_2 - \lambda_1 - \varpi_{1,2}$ (\third),
and the 4:1 MMR angles $\Theta_{1,2}^{4:1} = 4\lambda_2 - \lambda_1 - 3\varpi_{1,2}$ (bottom).
The MMR angle variables correspond to an inner mean motion resonance \citep{2018MNRAS.477.3383F}.
The two-dimensional surface density distribution of model \model{M9-3} is displayed in
Fig.~\ref{fig:2d_surface_density}, where panels a\,-\,f show the disk during three different stages:
prior (a,b), during (c,d), and after (e,f) the migration jump.
The times of the individual snapshots are indicated in Fig.~\ref{fig:migration-sample-zoomin} by the vertical lines.
Panel a of Fig.~\ref{fig:2d_surface_density}
shows the disk at that point in time when disk feedback is turned on and the planets are allowed to migrate.

In the following paragraphs we examine the migration jump process more closely by analysing the behaviour around
the time of each of the six snapshots of model \model{M9-3}.

\begin{enumerate}[label=\alph*)]
  \item Prior to the migration jump, both planets migrate outward in 2:1 MMR.
        During this time, the eccentricity of the inner planet, $e_\text{in}$, grows up to 0.15 while the
        outer planet's eccentricity, $e_\text{out}$, fluctuates up to 0.125.
        The increase in eccentricity comes from the interaction of the planets 
        with the vortex that is formed outside the common gap.
        Faint spiral arms are visible in the common gap during this epoch (see panel b in Fig.~\ref{fig:2d_surface_density}).

  \item At 64\,kyr, $e_\text{in}$ drops significantly while $e_\text{out}$ rises up to 0.2 through  and the 2:1 MMR is broken.
        Due to its eccentric orbit, the outer planet comes close to the inner edge of the outer disk
        (see panel b of Fig.~\ref{fig:2d_surface_density}).
        This is close enough to sufficiently enhance the mass flow across the planet's orbit and producing co-orbital torques.
        These torques are positive, since gas with higher angular momentum flows inwards where it has a lower angular momentum.
        The difference in angular momentum is deposited onto the planet causing a positive torque.
        What follows is a phase of type III rapid outward migration \citep{2008MNRAS.387.1063P}
        and the outer planet moves out from 40\,au at 65\,kyr to 72\,au at 70\,kyr, covering 32\,au in only 5\,kyr.
        More complex structures can be seen in the disk (see panel c in Fig.~\ref{fig:2d_surface_density})
        which are the result of overlapping spiral arms caused by both planets and the outer planet's gap opening.
        During this time $e_\text{out}$ relaxes back to low values.

  \item The fast outward migration stops at the location of the 4:1\,MMR with the inner planet (see panel three
        in Fig.~\ref{fig:migration-sample-zoomin} where the resonant angles $\Theta_{1,2}^{4:1}$ are displayed).
        The two planets remain in the 4:1\,MMR for about 4\,kyr at $t \sim 70$\,kyr, where the outer planet remains
        around $75$\,au.

  \item Afterwards the 4:1 resonance is broken and the outer planet migrates back inward by 22\,au within 3.5\,kyr.

  \item It ends up back in 2:1 MMR with the inner planet.
        The inner planet migrates outwards during the jump since the positive torque from the inner disk is dominating
        over the negative contribution from the outside which is weakened due to the common gap.
\end{enumerate}

Migration jumps occur in a variety of models but details such as its distance or duration 
of specific sub-processes vary from model to model.
The location where the jump stops is not necessarily at the 4:1 period commensurability for all simulations, 
as can be seen for example in Fig.~\ref{fig:migration-irradiation} which also shows period ratios close to 4.5 and 5.
In general the location can be expected to be determined by the interplay of N-body dynamics (the resonances),
gas dynamics, and the rearrangement of gas density around the outer planet.

Migration jumps only happen in the models in which migration is directed outward, a vortex forms outside the common gap, and in which the disk mass is high enough.

From these two observations we can draw two criteria that must be met for migration jumps.
First, the planet mass ratio must be such that the pair of planets migrates outwards.
This can be the case when the inner planet is more massive than the outer one.
Second, the disk mass needs to be high enough to facilitate fast enough outward migration and to allow for the formation
of a significantly massive vortex which in turn leads to large enough eccentricities for the outer planet which are needed
to trigger type III rapid outward migration.
For disk masses too low the system only migrates outward smoothly and no vortex forms
(see also the low disk mass results for model \model{PDS70 IRR M/5}, Fig.~\ref{fig:migration-PDS70}).

\subsection{Orbit swaps\label{sec:orbit_swaps}}

In addition to migration jumps, some models show other features as well. For example, the planets in model \model{M9-4.5}
swap their orbits in close succession just before a migration jump
(Fig.~\ref{fig:migration-showcase} at 110\,kyr).
The orbital eccentricities reach very high values during the swapping process,
e.g. $e_\text{out}=0.58$ in model \model{M9-4.5}.
The accreting model, \model{A0.01}, also experiences a double swap (Fig.~\ref{fig:accretion-accreting} at 90\,kyr).

A model with different resolution, \text{L}, shows a single orbit swap (see Fig.~\ref{fig:orbit-swap}).
As a consequence,
the inner planet is then less massive than the outer one and the migration direction changes from outward to inward.
Just after the single orbit swap, the outer, more massive planet undergoes a small migration jump (5\,au).

The models with self-gravity enabled also show orbit swaps.
Model \model{SG} shows a single orbit swap following two migration jumps and model \model{SG IRR}
shows a migration jump followed by a double orbit swap and a subsequent single orbit swap.
In each of our non-self-gravitating simulations where an orbit swap (single or double) happened,
a migration jump followed after a few more orbits.

Events like orbit swaps have already been observed in the literature,
especially in planetary systems with even more than two massive planets
\citep[e.g.][]{2010A&A...514L...4M,2011ApJ...729...47Z}.

\subsection{Vortex outside the gap}
\label{sec:vortex}

During the regular outward migration of the planet pair, an overdensity is formed just outside the common gap, visible
as the banana-shaped high density structure just beyond the outer planet in
panel b of Fig. \ref{fig:2d_surface_density}. 

This is a sign of vortex formation by embedded planets in disks, which are known to occur for either massive planets 
and/or low viscosity disks \citep[e.g.][]{2003ApJ...596L..91K,2013A&A...553L...3A}.
Fig.~\ref{fig:streamlines} zooms in to the overdensity and shows the surface density and streamlines in a system 
corotating with the overdensity.
The eye of the vortex is visible in the upper right quadrant.
Vortensity and surface density are analysed in more detail in Appendix~\ref{sec:app_vortex}, which
confirms that the overdensity is indeed a vortex.
In all our simulations in which a migration jump happened we also observed a vortex forming outside the common gap.

\subsection{Accretion onto star and planets}
\label{sec:accretion}

Figure~\ref{fig:accretion-accreting} shows from top to bottom panel: semi-major axis of both planets (top),
eccentricity of inner and outer planet, $e_\text{in/out}$ (\second, \third),
planetary mass accretion rate onto inner and outer planet, $\dot{M}_\text{in/out}$ (\fourth,\fifth),
and the mass accretion rate onto the star smoothed with a moving average of
length $1.185\,\text{kyr}$, $\dot{M}_*$ (bottom).

The time evolution of different masses is displayed in Fig.~\ref{fig:mass}.
It shows from top to bottom panel:
the mass accreted onto the inner (top) and outer (\second) planet, $M_\text{acc,pl}$,
the mass which leaves the disk through the inner domain, $M_\text{acc, *}$ (\third),  
(i.e. the mass that would be accreted onto the star, but which is not added to it in the simulations), 
and the evolution of the total disk mass, $M_\text{disk}$ (bottom).
The latter value also includes the mass leaving the outer domain.

\subsubsection{Planetary accretion}

\begin{figure}[t]
  \centering
  \includegraphics[width=\linewidth]{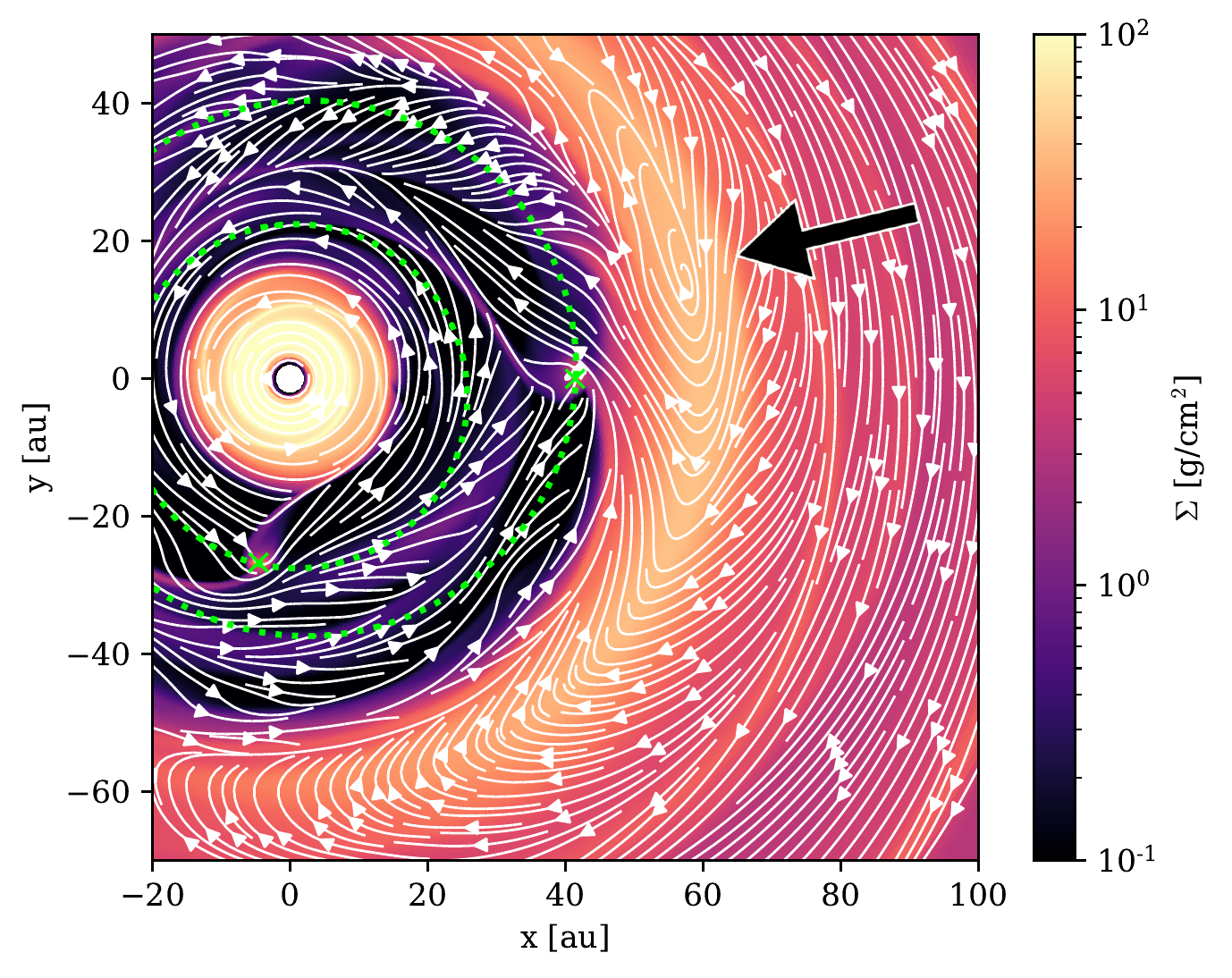}
  \caption{Zoom-in to panel b of Fig.\ \ref{fig:2d_surface_density} showing the surface density and velocity streamlines.
    The streamlines are computed in a frame corotating with the disk at 59\,au.
    The streamlines are closed in the region indicated by the black arrow,
    showing that the overdensity is a vortex.
    The orbits of the planets are indicated as the green dotted lines.
    \label{fig:streamlines}}
\end{figure}

Mass accretion onto the planets is turned on when the planets are released.
The accretion rates onto inner and outer planet, $\dot{M}_\text{in/out}$,
scale approximately linearly with $f_\text{acc}$, as one might have expected (see Eq.~\ref{eqn:mass_accretion_planet})
This behaviour holds during the outer planet's inward migration into 2:1 MMR
and during outward migration of the pair as long as the eccentricity of the outer planet stays below $e_\text{out}=0.1$.
Due to the larger relative size of the inner planet's Hill sphere ($R_\text{Hill} / a \propto M_\text{pl}^{1/3}$) and its 
smaller orbital period, $\dot{M}_\text{in}$ is larger than $\dot{M}_\text{out}$.
This is because the mass available for accretion by the inner planet is via the streamers within the common gap.
Then the accretion rate is determined by the ratio between a planet's region of influence ($R_\text{Hill}$) and the 
length of its orbit ($a$),
i.e. to how much of the mass around its orbit is accessible to it,
and by its orbital frequency, i.e. how often it encounters these streamers.
The larger accretion rate of the inner planet also shows
that the outer planet only receives a small fraction of the gas which otherwise travels through the common gap
and does not starve the accretion of the inner planet.
During these quieter times, the models behave very similar, independent of $f_\text{acc}$,
although the $f_\text{acc} = 0.1$ case shows the same events at a slightly earlier time.

\begin{figure}[t]
  \centering
  \includegraphics[width=\linewidth]{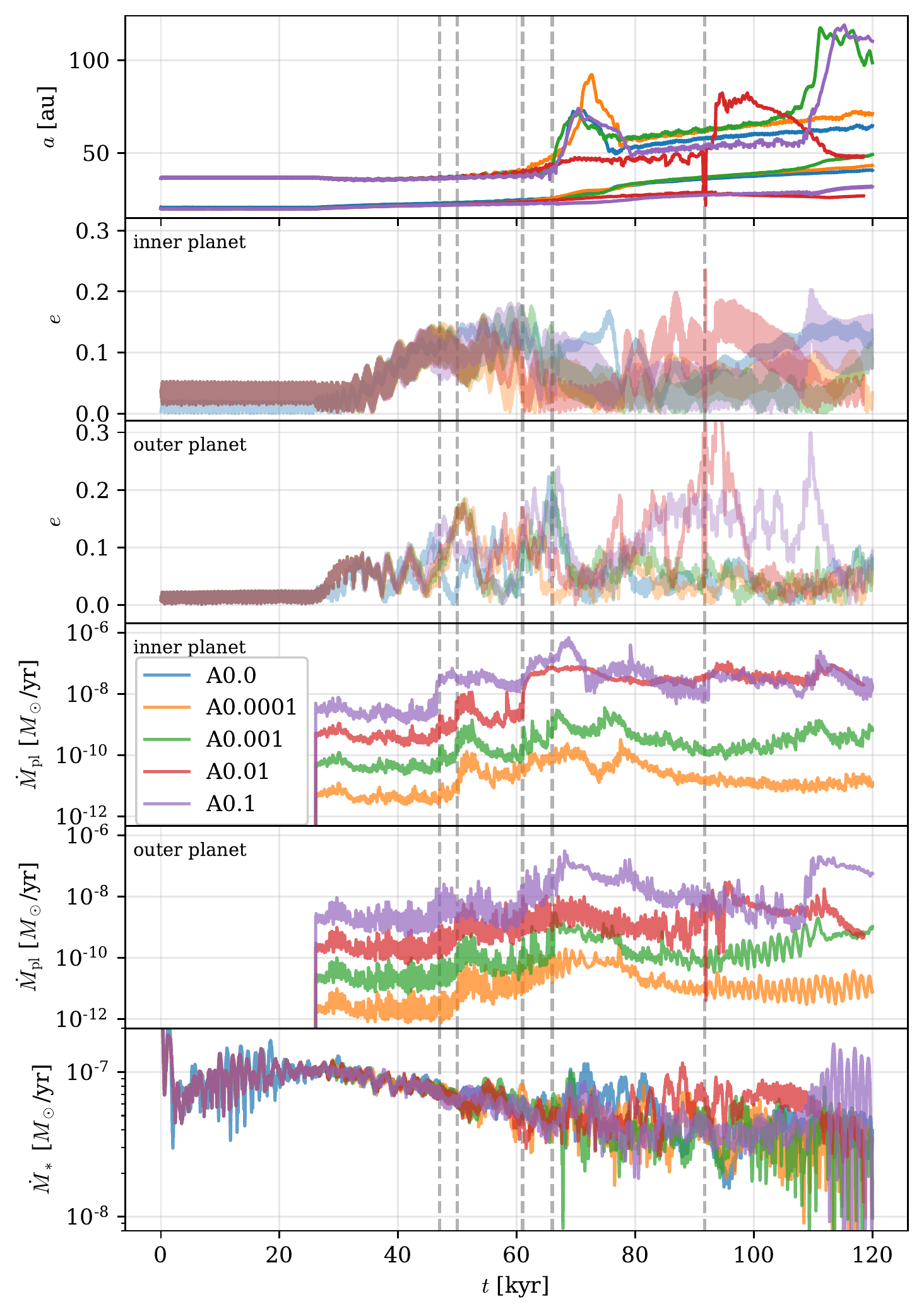}
  \caption{Evolution of accretion rates, migration, and eccentricities for the models with planetary accretion.
    The vertical dashed lines indicate events of interest and are referred to in Sect. \ref{sec:accretion}.
    The panels show semi-major axis of both planets (top), eccentricity of inner (\second) and outer planet (\third),
    planetary accretion rate of inner (\fourth) and outer planet (\fifth), and stellar mass accretion (bottom).
    Model \model{A0.0} is an alias for model \model{M9-3}.
    \label{fig:accretion-accreting}}
\end{figure}

Around $t=47\,\text{kyr}$ (\first\  vertical line in Fig.~\ref{fig:accretion-accreting})
when $e_\text{out}$ rises to values above 0.1, $\dot{M}_\text{in/out}$ increases as well.
This can be explained by the following argument.
When the outer planet comes close to the outer gap edge at apastron, it 'shovels' in disk material into the common gap
(see panel b in Fig.~\ref{fig:2d_surface_density} for the emerging structures).
Thus, there is more gas available inside the planets' Hill spheres to be accreted.
This way, outward migration with pumping of eccentricities can enhance planetary accretion.
Variabilities of $\dot{M}_\text{in/out}$ occur on timescale of around two times the outer orbits period.
$\dot{M}_\text{in}$ follows the same trends as $\dot{M}_\text{out}$ with a delay of around one outer period
(approx. 200\,yr),
due to the finite gap crossing time.

At around $t=50\,\text{kyr}$ (\second\  vertical line in Fig.~\ref{fig:accretion-accreting}),
$\dot{M}_\text{in/out}$ reach their highest values prior to the migration jumps before they decrease when $e_\text{out}$
relaxes to lower values.
During the smooth phase of outward migration, $\dot{M}_\text{in/out}$ can be increased by a factor of 10-20 compared to
inward migration.

Just before the onset of a migration jump (\third\  vertical line in Fig.~\ref{fig:accretion-accreting}),
$e_\text{out}$ and $\dot{M}_\text{in/out}$ rise again.
Models \model{A0.0001}, \model{A0.001}, and \model{A0.1} show a migration jump soon after, at
around $t\sim 66\,\text{kyr}$.
During the jump (\fourth\  vertical line in Fig.~\ref{fig:accretion-accreting}),
$\dot{M}_\text{out}$ is comparable to $\dot{M}_\text{in}$ because the outer planet moves through the disk and has a
higher density of gas inside its Hill sphere.
$\dot{M}_\text{in}$ rises as well, as significant amounts of gas are scattered inward during the jump.
This can result in a 50-100 fold increase compared to the values during inward migration.

A notable exception to the described behaviour is model \model{A0.01}.
There, the first migration jump fails which gives rise to another distinct behaviour.
Similar to the other models, the outer planet increases the mass flow through the gap by 'shovelling' in gas.
However, unlike the other models, it does not embark on a jump and continues to supply material to the inner planet.
This causes the inner planets Hill sphere's gas density to rise and $\dot{M}_\text{in}$ to rise tenfold,
matching the rates of the \model{A0.1} model.
At $t=91.7\,\text{kyr}$ (\fifth\ vertical line in Fig.~\ref{fig:accretion-accreting}),
$\dot{M}_\text{out}$ on \model{A0.01} matches the values of \model{A0.1} as well,
when it finally embarks on a jump and its eccentricity has risen to $e_\text{out} = 0.2$.

Depending on the efficiency of planetary accretion, $f_\text{acc}$, the changes in planet mass can be substantial,
as seen in the first two panels of Fig.~\ref{fig:mass}.
The most extreme case is the outer planet of model \model{A0.1} which has almost doubled its mass at the end of the simulation.

\subsubsection{Stellar Accretion}

\begin{figure}[t]
  \centering
  \includegraphics[width=\linewidth]{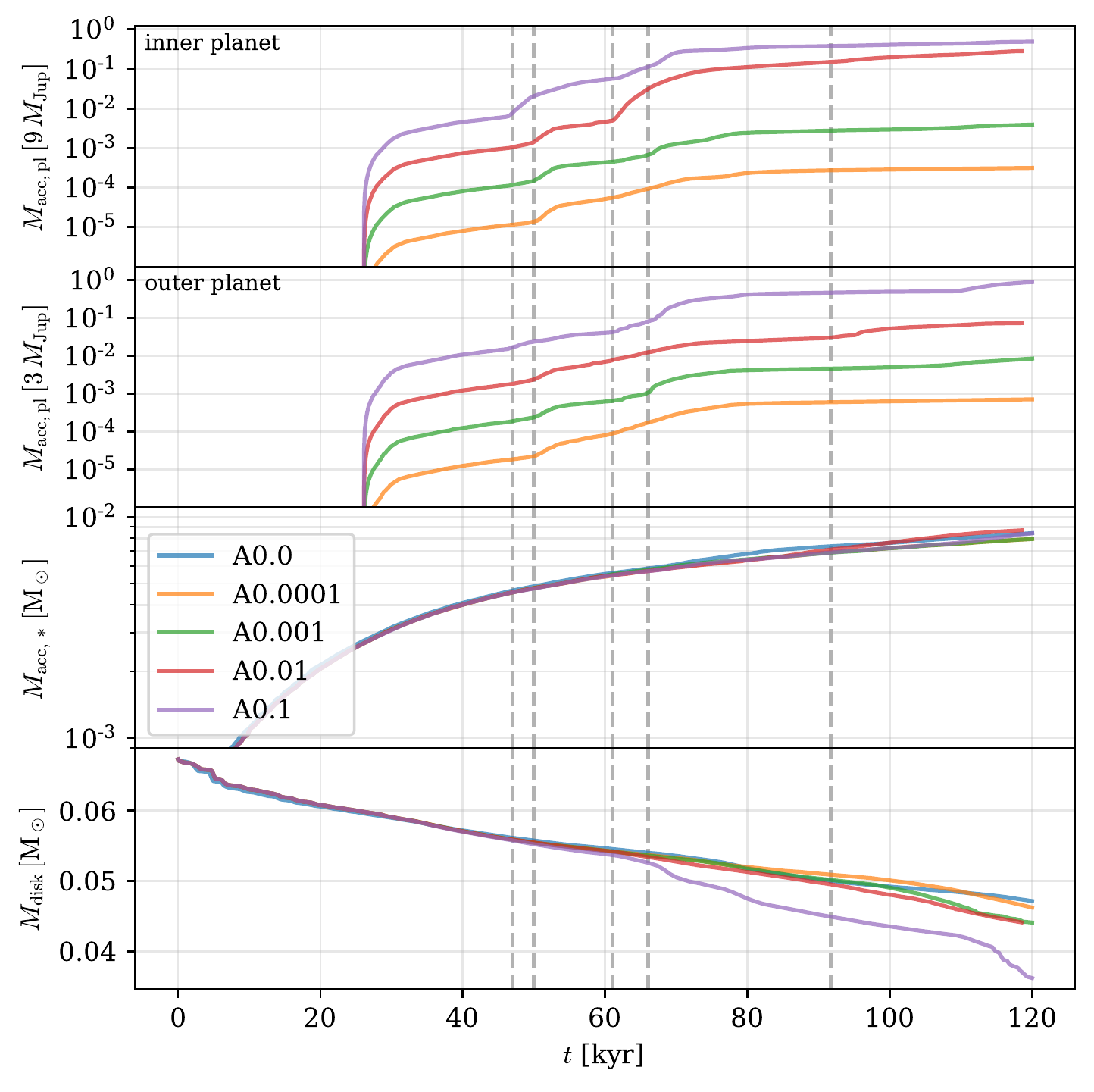}
  \caption{Evolution of the mass accreted onto the planets (top and \second panel), the star (\third),
  and the total disk mass (bottom) for the models with planet accretion.
  The values are in units of the initial mass of the respective object for the accreted masses and in stellar masses
  for the disk.
  The dashed vertical lines correspond to the ones in Fig.~\ref{fig:accretion-accreting}.
  \label{fig:mass}
  }
\end{figure}

Mass accretion onto the star, $\dot{M}_*$, is calculated by summing up the mass that leaves the inner boundary
over the output interval of 11.8\,yr.
It does not show a dependence on the efficiency of planetary accretion, $f_\text{acc}$.
The mass lost through the inner boundary is not added to the primary star since its contribution
to the total stellar mass over the time span of the simulations would be smaller than 1 per cent 
(see Fig.~\ref{fig:mass} panel three) and can be neglected.
For a theoretical estimate of the stellar mass accretion rate we can use the viscous mass accretion rate of the
unperturbed disk at the inner boundary ($\approx 2\,\mathrm{au}$) which is
$\dot{M}_\text{disk} = 3\pi\Sigma\nu = 5.3\times10^{-9}\,\text{M}_\odot/\text{yr}$.
In the simulations the mass accretion rates onto the star through the inner boundary fluctuate
between $10^{-8}$ to $10^{-7}\,\text{M}_\odot/\text{yr}$. 
One should keep in mind that the theoretical estimate does not take into account the presence of the massive 
embedded planets which can be expected to alter the disk's dynamics substantially.

This discrepancy could be explained by the inner outflow boundary and changes in the disk structure.
Because there is no pressure support from the inside, there is a high mass flow in radial direction.
An additional contribution arises, when the disk becomes eccentric in its inner region.
The simulations show gas eccentricities between 0.1 and 0.4 in the inner 10\,au of the domain, which are a result of 
the gravitational interaction of the planets and the disk.
Since the boundary is perfectly circular by design, any gas that is on an eccentric orbit which overlaps with the boundary is lost through the boundary
because it cannot reenter although its orbit would bring it back into the domain.
Therefore, the accretion rate onto the star, $\dot{M}_*$, can only be seen as an upper limit.

Models \model{VB-P} and \model{VB5-P}, which employ a viscous boundary to avoid abnormally high mass flow through the inner
boundary, show mass accretion rates at a similar level
with values around $\dot{M}_* \sim 2 \times 10^{-8} \,\text{M}_\odot/\text{yr}$ for both choices of $v_\text{in}$
($\beta = 1$ or 5 in Eq.~\eqref{eqn:viscous_boundary}).
This means, that the inner disks' surface density must be enhanced compared to the initial condition, which is indeed the case
for simulations showing outward migration.
This result can be taken as an indication, that the outflow boundary in our simulations does not strongly overestimate the mass accretion
through the inner boundary.
On the contrary, in our simulations, the stellar accretion rate seems to be determined by the amount of gas which can be supplied from further out in the disk.

\begin{figure}[t]
  \centering
  \includegraphics[width=\linewidth]{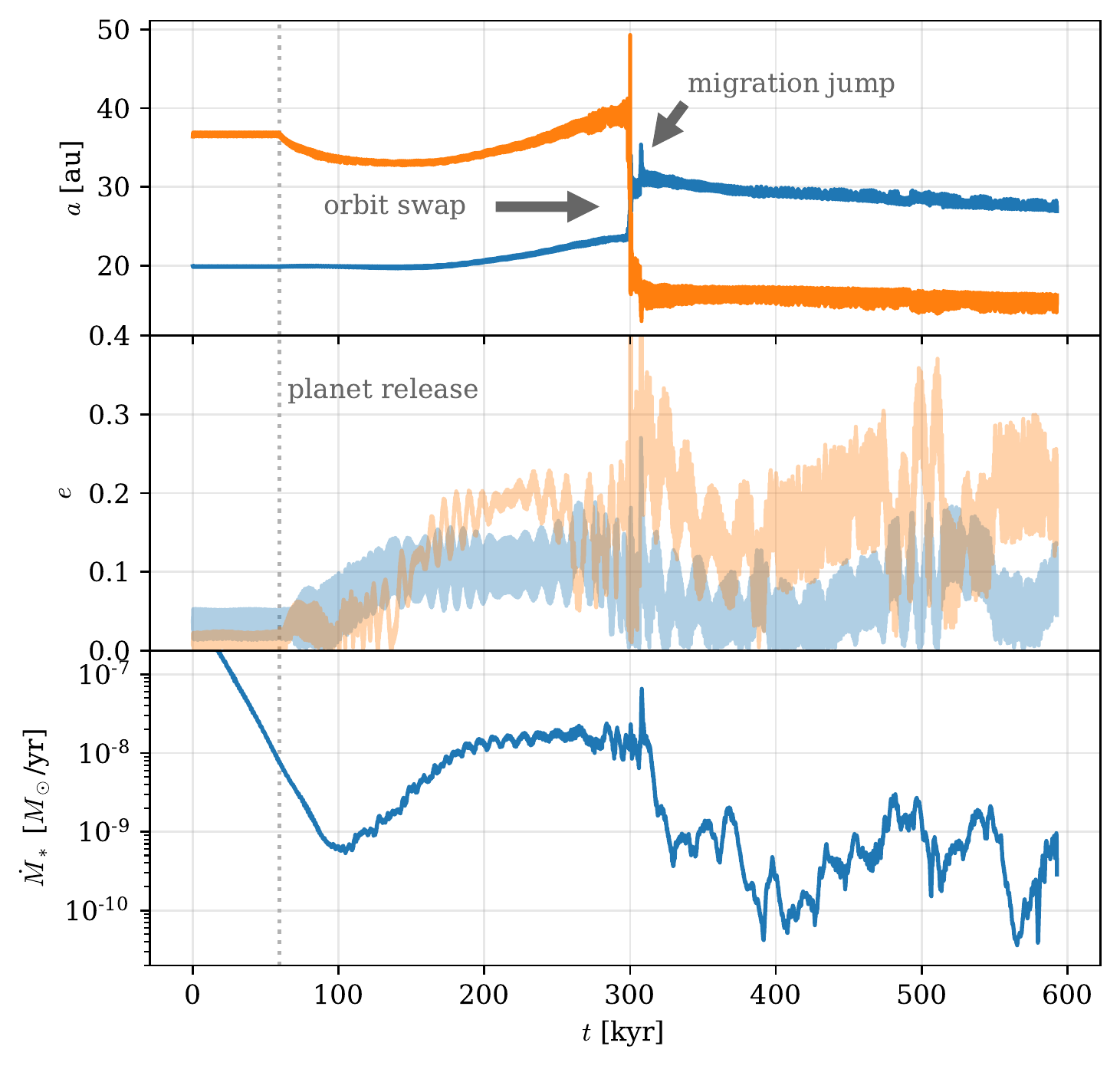}
  \caption{Migration and accretion for model \model{L} showing the orbit swap and how $\dot{M}_*$ depends
  on the direction of migration.
  The panels show (top) semi-major axis, $a$, (middle) eccentricity, $e$,  and (bottom) migration onto the star $\dot{M}_*$.
  The vertical dashed line indicates the time when planets are allowed to migrate.
  \label{fig:orbit-swap}}
\end{figure}

A special behaviour was observed in model \model{L} which exhibits a single orbit swap, as explained in Sect.~\ref{sec:orbit_swaps}.
Because migration changes from outward to inward in an otherwise unchanged  disk,
this model gives us the opportunity to study how $\dot{M}_*$ depends
on the direction of migration and the ordering of the planets.
Figure~\ref{fig:orbit-swap} shows migration and eccentricities of both planets (top and middle panel)
and $\dot{M}_*$ smoothed with a moving average of length $1.185\,\text{kyr}$ (bottom panel) for model \model{L}.
The small $\dot{M}_*$ at $t=100\,\text{kyr}$ is due to the initialization phase when the inner disk is mostly depleted
by accretion through the inner boundary.
In the time after planet release, the inner disks recovers parts of its mass and reaches up to a tenth of its
initial value of $\Sigma$ at $t=150\,\text{kyr}$ and a fourth at $t=250\,\text{kyr}$ by mass transfer through the gap.
This is enough to start outward migration.
In the 120\,kyr of outward migration leading up to the migration jump,
accretion is enhanced at $\dot{M}_* \approx 2\times 10^{-8}\,\text{M}_\odot/\text{yr}$
(see bottom panel of Fig.~\ref{fig:orbit-swap}).
After the planets swap orbits at 300\,kyr, migration changes its direction to inward and accretion deceases to
values $\dot{M}_*<10^{-9}\,\text{M}_\odot/\text{yr}$.
The eccentricity of the massive planet and the value of $\dot{M}_*$ follow similar trends
(see Fig.~\ref{fig:orbit-swap}).
However, both values are not proportional which can be seen by comparing the values at 200\,kyr and 520\,kyr where
the massive planet's eccentricity is at a value of 0.1, but $\dot{M}_*$ is at least 10 times smaller at the later time.
This rules out that the increase in $\dot{M}_*$ during outward migration is a result of the outflow boundary
effect explained in the last paragraph.
During the event of the orbit swap, there is no abnormally high mass loss of the disk.
The disk mass in the whole domain only changes from 0.145 to 0.14\,$\text{M}_\odot$ after the planets are released so
the change in $\dot{M}_*$ cannot be explained by the disk suddenly loosing most of its mass.
Thus, $\dot{M}_*$ depends on the direction of migration and it is enhanced for outward directed migration.
The model also shows that mass can be transported through a common planet gap at a substantial rate.

Summarizing the results, we can report the following findings:
\begin{enumerate}
  \item Mass accretion through the disk can be sustained even in the presence of a large common gap carved by a pair of
        massive planets.
  \item Planetary accretion can be substantially increased when a pair of planets is migrating outward in 2:1 MMR where the
        increase in mass accretion can be 10-20 fold compared to inward migration.
  \item During extreme events like a migration jump, when a planet travels through previously unperturbed disk regions,
        the increase can be even higher with values increased 50-100 fold compared to inward migration.
  \item Migration rates seem not to be affected by the choice of $f_\text{acc}$ during times of smooth outward migration.
        However, the accretion efficiency can change the type of occurring events and the timing when they happen.
  \item Mass accretion rates onto the star seem not to be affected by the choice of $f_\text{acc}$.
\end{enumerate}

\subsection{Planet ejection}
\label{sec:planet_ejection}

In some models, the planetary system is ejected either during a first or second migration jump.
The outer planet is ejected first and the inner planet is ejected 2-3\,kyr later.
The models affected are \model{VB}, \model{VB5}, \model{WD}, and \model{WDR}.
All of them use a disk (hydro simulation) centered around the center of mass of the star and the two planets
and have an inner boundary condition different from outflow.
The affected inner boundary conditions are the viscous boundary with $v_\text{in} = v_\text{visc}$ and
$v_\text{in} = 5 v_\text{visc}$, outflow with an additional wave damping zone, as well as reflective with an additional
wave damping zone.
We repeated all the affected models with the disk (hydro simulation) centered on the primary star
(models \model{VB-P}, \model{VB5-P}, \model{WD-P}, and \model{WDR-P}).
In the primary frame, no ejection occurred during the simulation time which was chosen at least 50\,kyr longer than
the time when the ejection happened in the respective center of mass model.

In all the models, the inner boundary is located at 2.08\,au and the wave damping zone stretches from the inner boundary
to 4.16\,au.
This comparison shows, that the inner boundary can be crucial for determining the fate of an embedded planetary system,
even if planets are further out at around 50\,au.

All ejection occurred in models where the primary star was moving with respect to the boundary.
In the center of mass frame the primary moves up to 0.4\,au, 20 per cent of the inner boundary radius.
This equivalently means, that the boundary is moving 0.4\,au with respect to the star.

The difference is not only a numerical artefact, but the physical boundary condition is different for the models in the
center of mass frame compared to the primary frame.
Using the viscous boundary condition in the primary frame for example, the radial speed at a fixed radius from the star is
set to the viscous speed.
This is physically motivated choice.
Since the hydro grid is centered on the primary, this boundary condition can be implemented in a simple way by setting
the velocity at the inner boundary, which has a constant distance to the star in any direction.
Using the same implementation in a center of mass frame, like we did in our models, the distance at which the radial
velocity is set to the viscous speed is different depending on the azimuthal direction and is even varying over time.
Thus, the resulting boundary condition is physically different from the case of the primary frame and is not physically
plausible any more.

In all cases, except the zero-gradient boundary condition case, the unphysical boundary condition
leads to strong perturbations close to the inner boundary which, at some point in time,
can not be resolved by the hydro simulation any more.
In the resulting numerical instability the perturbations grow unbound and destroy the disk and 
coincidently also eject the planets without any close encounter or instability in the N-body system.

\subsection{Final location of outward migration}
\label{sec:final_location}

Most models were run for a simulation time of $120\,\text{kyr}$.
At this point in time, outward migration did not halt in any of the models.
Some models were integrated for a longer time, e.g. the \model{M9-3} model in which the outer planet reached 133\,au at
$t=226\,\text{kyr}$.
In this specific case, the outer gap edge is located at 162\,au and the outer disk has two orders of magnitude lower
surface density compared to the inner disk.
As a result, the negative torque contribution from the outer disk gets diminished and outward migration continues.
Our models suggest, that in the scenario studied here, the case of outward migration of a pair of planets in a
disk of sufficiently high mass, the final location of the planet pair will be near the outer edge of the disk.

\section{Observability}
\label{sec:observability}

\begin{figure*}[t]
  \centering
  \includegraphics[width=\linewidth]{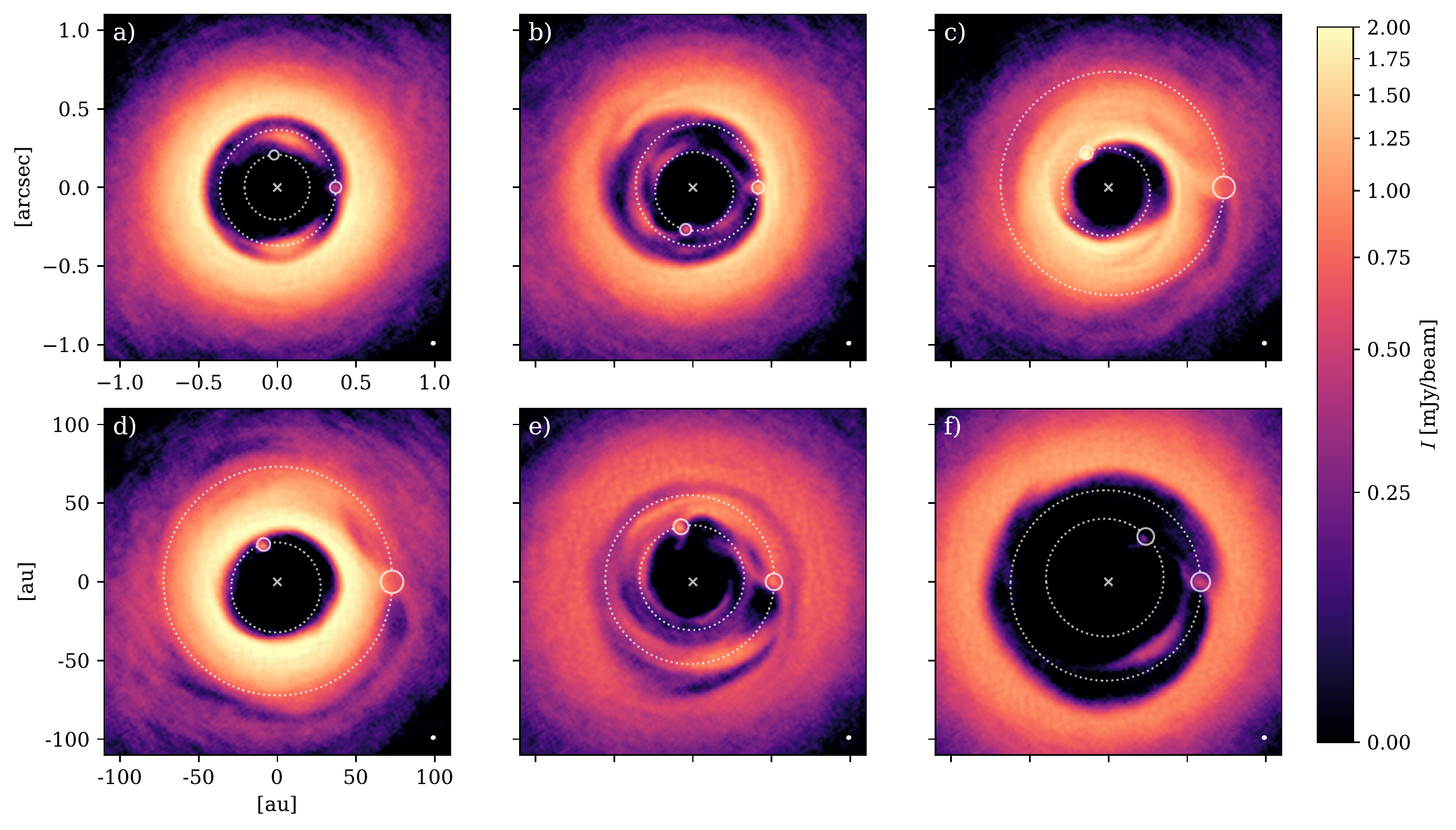}
  \caption{\label{fig:synth_obs_montage}
    Synthetic ALMA observations at $855\,\mu$m of the disk in model \model{M9-3} (assuming a distance of 100\,pc)
    at different times: prior (a and b), during (c and d), and after (e and f) a migration jump.
    The panels coincide with the ones in Fig.~\ref{fig:2d_surface_density} but they show the
    intensity from simulated observations instead of surface density
    and are zoomed in to the inner $\pm 100\,\text{au}$ of the disk.
    Coordinate ticks are the same in all panels and values are given in arcseconds in the top left panel
    and the corresponding values in au are shown in the bottom left panel.
    The ellipse in the bottom right corner of each panel indicates the beam size of $33\times 30\,\text{mas}$.
    The location of the star is indicated by the small $\times$ in the center.
    The current orbits of the two planets are marked by the dotted white ellipses and the planetary Hill spheres are
    indicated by the small circles.
  }
\end{figure*}

To evaluate the possibility, whether the effects of migration jumps could be observed in real systems,
we performed synthetic observations.
Though the timescale of the processes discussed here is quite short, some 10s of kyr if we also include the
time the structural changes last in the disk (see panel e and f of Fig.~\ref{fig:synth_obs_montage}), 
the synthetic observations might be applicable to systems in which we directly observe embedded planets, such as PDS~70.
These observations are, after all, just snapshots in time of the real system.
With a disk lifetime of the order of roughly 1\,Myr, the migration jump timescale of 10\,kyr amount to a significant 
fraction in the per cent region.
Considering the ever-increasing number of detected (proto-)planetary systems and the apparent richness in substructure
therein \citep{2018ApJ...869L..41A}, the following synthetic images might provide an explanation for a subset of 
future disk observations.

Our synthetic images were produced by calculating the thermal emission using RADMC3D
\citep{2012ascl.soft02015D}.
This emission was then postprocessed using the CASA package \citep{2007ASPC..376..127M} to simulate the instrumental effects produced by ALMA.

The resulting synthetic observations of the \model{M9-3} model are shown in Fig.~\ref{fig:synth_obs_montage} and
can be directly compared to the plots of surface density in Fig.~\ref{fig:2d_surface_density}.
The simulation snapshots are the same in the respective panels.

\subsection{Radiative transfer model} \label{sec:radtrans}

In order to compute the thermal dust emission for the individual snapshots of the \model{M9-3} model
we convert the gas surface density profiles to a three-dimensional dust density model which serves as input
for the radiative transfer calculations with RADMC3D.
Please keep in mind that, in principle, dust follows its own dynamics and is only partially coupled to the gas.
Our synthetic images are therefore only an approximation to what would be the result of the actual dust distribution 
taking into account the proper gas-dust interaction.
In the model we employ 8 dust species with a combined dust to gas mass ratio of $10^{-2}$.
The dust grain sizes are logarithmically evenly spaced, ranging from $0.1 \, \mu \mathrm{m}$ to 1 mm.
The number density size distribution of the dust grains follow the MRN distribution $n(a) \propto a^{-3.5}$,
where $a$ is the grain size.

Dust settling towards the mid-plane is considered following the diffusion model of \cite{1995Icar..114..237D} with
dust vertical scale height
\begin{equation}
  H_\mathrm{d} = \sqrt{\frac{\alpha}{\alpha + \mathrm{St}}} H \,,
\end{equation}
where $\mathrm{St}$ is the local Stokes number
\begin{equation}
  \mathrm{St} = t_\mathrm{stop} \, \Omega_\mathrm{K} = \frac{\pi}{2} \frac{a \, \rho_\mathrm{d}}{\Sigma} \,,
\end{equation}
with the grain density $\rho_\mathrm{d} = 3.0 \, \mathrm{g} / \mathrm{cm}^3$.
We assume a Schmidt number of one. Furthermore, the aspect ratio is assumed to be flared with radius,
i.e. $H/r$ scales as $(r / r_0)^\gamma$ where $\gamma$ is the flaring index.
In all the synthetic images a flaring index of $\gamma = 0.25$ was chosen.

For the extension in the polar direction 32 cells are equally spaced in their angular extent between
$\theta_\mathrm{lim} = \pi/2 \pm 0.3$ resulting in a maximal spacial extent of $z_\text{lim} = r\,\sin(\pm 0.3)$.
The vertical disk density profile is assumed to be isothermal and the conversion from the surface density
to the local volume density is calculated as follows:
\begin{align}
  \rho_\mathrm{cell} = \frac{\Sigma}{\sqrt{2\pi} H_\mathrm{d}} & \cdot \mathrm{erf}^{-1} \left( \frac{z_\mathrm{lim}}{\sqrt{2} H_\mathrm{d}} \right)
  \\ &\cdot \frac{\pi}{2} H_\mathrm{d} \frac{\ \left[ \mathrm{erf}\left( \frac{z_+}{\sqrt{2} H_\mathrm{d}} \right) - \mathrm{erf}\left( \frac{z_-}{\sqrt{2} H_\mathrm{d}} \right)  \right]}{z_+ - z_-} \,.
\end{align}
The error function term is a correction for the limited domain extent in the vertical direction that would otherwise lead to an underestimation of the total dust mass.
Similarly, the second correction term accounts for the finite vertical resolution, especially important for thin dust layers with strong settling towards the mid-plane.
The coordinates $z_+$ and $z_-$ are the cell interface locations in polar direction along the numerical grid.
For each grain size bin and a wavelength of 855 $\mu \mathrm{m}$ the corresponding dust opacities were taken 
from the \texttt{dsharp\_opac} package which provides the opacities presented in \cite{2018ApJ...869L..45B}. 
These opacities are based on a mixture of water ice, silicate, troilite, and refractory organic material.
In the RADMC3D model the central star is assumed to have solar properties with effective temperature of 6000 K 
at a distance of 100 pc. 
For the thermal Monte-Carlo simulation a number of $n_\mathrm{phot} = 10^8$ photon packages and for the 
image reconstruction $n_\mathrm{phot\_scat}=10^7$ photon packages were used.
Scattering of photons is assumed to be isotropic.

As a result of our assumption that gas and dust share the same spacial distribution
the surface density of dust in the inner disk is very high.
This leads to the formation of a hot dusty wall on the inside that prevents us from seeing the features in the 
outer disk, close to the planets.
Therefore, we reduce the dust density for the inner region by a factor of $10^{-5}$ in the radiative transfer model.
The cutoff radius for this reduction in dust density is set to 23 au for the first five snapshots in 
Fig. \ref{fig:synth_obs_montage} whereas it is extended to 26 au in Fig. \ref{fig:synth_obs_montage}.
Since dust is expected to drift within the inner disk, isolated by the two planets which leads to a reduction
in dust density there, this numerically motivated measure has also some physical foundation.
However, full consideration of the gas-dust interaction would be necessary to clarify 
how much the effects differ in strength.
Here, the reduction is a numerical measure to aid the visualization.

RADMC3D uses different and arguably more realistic opacities to calculate the dust temperatures compared to the
opacities used in the hydro simulations.
Therefore, the dust temperatures are different from the temperatures 
in the hydro simulation.
Compared to the gas temperature in the \model{IRR} model (Fig.~\ref{fig:hoverr-irradiation}) and 
depending on the dust grain size,
the resulting dust temperatures are 3-5 times higher in the gap region, 
which is directly illuminated due to the reduction of the inner disk's dust density,
2-3 times higher in the outer disk close to the gap and approximately the same value at 100\,au.

\subsection{Synthetic imaging}
\label{sec:synthobs}

For simulating the detectability of the various features present in the model we use the task \texttt{simalma}
from the CASA-5.6.1 software.
A combination of the antenna configurations \texttt{alma.cycle5.8} and \texttt{alma.cycle5.5} was chosen.
The simulated observation time for configuration 8 is four hours while the more compact configuration is integrated over
a reduced time with a factor of 0.22. \\
A simple auto-cleaning procedure was applied to reduce the artificial artefacts from the incomplete uv-coverage.
In the scope of this paper the prescription is sufficient for estimating the observability of the features of interest.
Consistent with the radiative transfer model, the observed wavelength is simulated to be at $855 \, \mu\mathrm{m}$,
corresponding ALMA band 7. \\
The resulting beam size is $33 \times 30 \, \mathrm{mas}$. For model \model{M9-3} the rms noise intensity ranges
from $40 \, \mu\mathrm{Jy/beam}$ to $70 \, \mu\mathrm{Jy/beam}$.
The synthetic images are shown in Fig. \ref{fig:synth_obs_montage}.

In the rest of this section, we will describe the different features that are visible in the synthetic images.
The mentioning of a panel will refer to Fig.~\ref{fig:synth_obs_montage}, if not stated otherwise.

\subsection{Features in synthetic images}

Our synthetic images show a \textbf{large inner hole} in the disk which is growing over time.
It can appear as a circular hole at some points in time (panels a and f)
or as an \textbf{eccentric hole} at other times (panels b, c, and d).
Since we removed the dust artificially,
the inner hole might actually be filled with dust partially and still be showing a visible inner disk.
Since our cut-off radius is 26\,au at maximum,
such a system with inner disk would show a \textbf{large gap} ranging from about 20\,au in size for
panel b to around 75\,au for the disk in panel f.
For an example where only part of the dust is removed, see the PDS\,70 models in Fig.~\ref{fig:synthobs-PDS70}.

Our simulations produce a number of non-axisymmetric features in the synthetic observations.
In the following sections, we will investigate these in more detail.

Emission from the location of the planets is clearly visible in the synthetic observations.
Very localized, \textbf{point like} features around the planet are visible either from a single planet (panels a,c, and d)
or from both planets (panels b, e, and f).
They emerge due to mass accumulations in the planets' Hill spheres.
During a migration jump (panels c and d), the outer planet is deeply embedded in the disk, so no separate point like
emission is visible from the outer planet.

The \textbf{vortex} discussed in Sect.~\ref{sec:vortex} is visible also in the synthetic images.
It appears as a bright region on the right side of panel b, causing a visible non-axisymmetric feature.
Depending on the line of sight inclination of the disk and beam size, the asymmetry caused by the vortex can even be
enhanced (see panel b and d of Fig.~\ref{fig:synthobs-PDS70}).

\label{sec:spur}
When the outer or inner planet gets close to the gap edge, the outer part of the spiral arm can be visible in the
synthetic images, connecting the planet with the outer disk with a visible \textbf{spur}.
This is the case for panels b, e and f with the outer planet and panel c and e with the inner planet.

At the time when the planets are released,
there is still some material present in the $\text{L}_4$ and $\text{L}_5$ points of the outer planet.
This is a left over of the initialization process.
Although it is also visible in the synthetic observations (panel a),
it does not constitute a realistically observable feature but rather an artefact of the initialization
and the assumption that dust and gas share the same spacial distribution.
However, when the outer planet migrates back in during a migration jump,
a substantial amount material survives as a \textbf{mass accumulation in Lagrangian point $\text{L}_5$}
(trailing the planet).
The accumulation is clearly visible in panel e.
This feature can persist for more than $20\,\text{kyr}$ as its presence in panel f shows.

During times of high mass transfer through the common gap, surface density in the spiral arms in enhanced.
Panel b, c, and e show clear signs of \textbf{spiral arms} even in the synthetic images.
The spiral arms of both planets are periodically cut off by the passing of the other planet and locally merge with
parts of the other planets spiral arm.
This causes additional \textbf{arc} like features such as the ones in panel b and e.

Another \textbf{arc} like feature can be produced during the migration jump.
Panel c and d show a \textbf{void} behind the planet in the lower right quadrant at a radius of approximately 75\,au.
These voids are $30\deg$ (panel c) to $60\deg$ (panel d) large in azimuthal direction and are caused
by the gap being carved into the previously unperturbed disk during rapid outward migration
\citep{2008MNRAS.387.1063P}.

Just after the migration jump (panel e), multiple substructures can be seen inside the planets' orbits.
This shows, that more than one of the features presented above can exist in a disk at the same time.

\section{Modelling of a real system: PDS 70\label{sec:PDS70}}

PDS\,70a is a $5.4\pm1.0\,\text{Myr}$ old K7-type star with a mass of $0.76\pm0.02\,\text{M}_\odot$ and luminosity outflow
$L_* = 0.35\pm0.09\text{L}_\odot$ \citep{2018A&A...617L...2M}.
It is at a distance of $113.43\pm0.52\,\text{pc}$ \citep{2018A&A...616A...1G}.
Recently, it was found to host two giant planets,
PDS\,70b and PDS\,70c which were observed via direct imaging. Their orbits are
close to a 2:1 MMR with distances of about $a_\text{b} = 20.6\pm1.2\,\text{au}$ and
$a_\text{c} = 34.5\pm2\,\text{au}$ \citep{2018A&A...617A..44K, 2019NatAs...3..749H}.
The inner planet is believed to be more massive than the outer one while their mass estimates are still uncertain
at $M_\text{b} = 5-14\,M_\text{Jup}$ \citep{2018A&A...617A..44K} and $M_\text{c} = 4-12\,M_\text{Jup}$
\citep{2019NatAs...3..749H}.
These masses were estimated by comparing photometry of the sources to synthetic colours from planet evolution models.

Recently, \citet{2019ApJ...884L..41B} showed that ALMA dust continuum observations at 890\,$\mu$m can be convincingly
reproduced by a pair of outward migrating planets.
They performed 2D locally isothermal hydro simulations with a temperature profile obtained from radiative transfer
calculations, using a stellar mass of $0.85\,\text{M}_\odot$ and two planets close to 2:1 MMR with masses of
$M_\text{b} = 10\,M_\text{Jup}$ and $M_\text{c} = 2.5\,M_\text{Jup}$.
They found the system to be stable for 1\,Myr while smoothly migrating outward.

We ran additional simulations to model PDS\,70 in order to test whether migration jumps could occur in that system.
The PDS\,70 models differ only slightly from our standard \model{M9-3} setup, the stellar mass is lowered to
$M_* = 0.76 \text{M}_\odot$.
One PDS\,70 simulation uses the locally isothermal equation of state (\model{PDS70 ISO}).
A second and third one use the ideal equation of state with irradiation from the star, like the \model{IRR} model above,
with stellar luminosity $L_* = 0.35\pm0.09\,\text{L}_\odot$ \citep{2018A&A...617L...2M}.
The first two models use the standard $\Sigma(r)$ profile (\model{PDS70 ISO}, \model{PDS70 IRR})
and last one has a 5 times smaller $\Sigma$ (\model{PDS70 IRR M/5}).
All three models were integrated for $100\,\text{kr}$ and are included in Table~\ref{tab:simulation_parameters}.

Our models have a higher surface density compared to the ones in \citet{2019ApJ...884L..41B}.
At a distance of 40\,au this amounts to $\Sigma = 11.6 \,\text{g}/\text{cm}^2$ and $\Sigma = 2.4 \,\text{g}/\text{cm}^2$
(\model{PDS 70 IRR M/5}) in our models compared to their $\Sigma = 1 \,\text{g}/\text{cm}^2$.

\subsection{Dynamical results}

\begin{figure}[t]
  \centering
  \includegraphics[width=\linewidth]{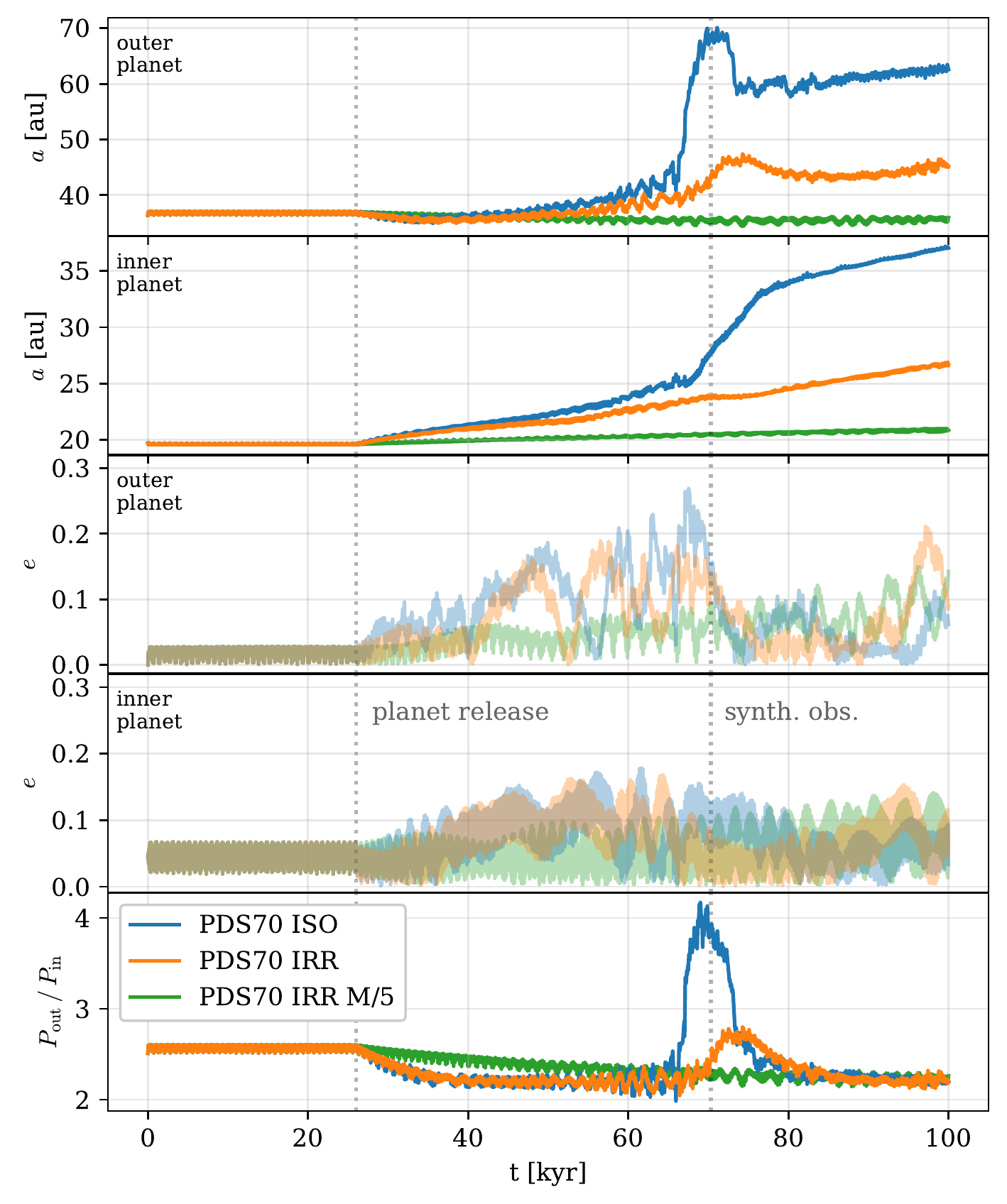}
  \caption{\label{fig:migration-PDS70}Migration history (as Fig.~\ref{fig:migration-showcase}) for the PDS\,70 models.
    A migration jump occurs for both equations of state in the case of a high surface density model
    (\model{PDS70 ISO} and \model{PDS70 IRR}).
    The model with a fifth of the surface density (\model{PDS70 IRR M/5}) shows only very moderate outward migration.
    The left vertical dotted line shows the time of planet release.
    The right vertical dotted line shows the time of the synthetic observations shown in Fig.~\ref{fig:synthobs-PDS70}.}
\end{figure}

The model with a lighter disk (\model{PDS70 IRR M/5}) shows no special events.
The inner planet migrates outward very slowly while the outer planet migrates inward until the system
is locked into 2:1 MMR around $t\approx 80\,\text{kyr}$ ($54\,\text{kyr}$ after planet release).
The system then migrates slowly outward together, maintaining the 2:1 resonance.
The inner planet moves less than $0.5\,\text{au}$ over the remaining $20\,\text{kyr}$.

Both models with high $\Sigma$ show one migration jump until the end of simulation at $100\,\text{kyr}$.
The outer planet first migrates inward for approximately 10\,kyr until the planets lock in 2:1 MMR.
Then rapid outward migration starts in both cases leading up to a migration jump 30\,kyr later.

In the \model{PDS70 ISO} case, the outer planet travels 10\,au during that time and a migration jump takes it out
to nearly 70\,au with a period commensurability of $P_\text{out}/P_\text{in} \approx 4$.
During the $16\,\text{kyr}$ duration of the migration jump, the inner planet travels out by $5\,\text{au}$,
because the negative torque contribution from the outside is missing.
Due to the fast outward migration of the inner planet, when the outer planet migrates back in from the jump,
the location of the 2:1 MMR is much further out.
When the system goes back into 2:1 MMR, the outer planet is at $60$\,au from where on outward migration continues.

The sibling simulation with a more realistic equation of state and irradiation from the star (\model{PDS70 IRR})
shows similar, though less extreme, effects.
The outer planet travelled out by $4\,\text{au}$ since the planets got locked into resonance $30\,\text{kyr}$ before
the migration jump is about to happen.
The smaller migration jump moves it out by $6\,\text{au}$ to $46\,\text{au}$,
where it has a period commensurability of $P_\text{out}/P_\text{in} \approx 2.75$.
Again, the migration jump takes around $16\,\text{kyr}$,
but the inner planet only migrates approximately 1\,au during that time.
Thus the location of 2:1 MMR for the outer planet is further in, at 44\,au, to where it migrates back over a
timespan of 6\,kyr, instead of jumping back, until it locks into 2:1 MMR again.
The different stopping location of the jump between models \model{PDS70} and \model{PDS70 IRR} does not seem to stem 
from a difference in aspect ratio.
In fact, the aspect ratio in model \model{PDS70 IRR} is nearly identical to the one in model \model{IRR},
which is displayed in Fig.~\ref{fig:hoverr-irradiation}, 
and is close to 0.05 around 40\,au, just like in the locally isothermal model.

Along with the migration jumps, both models with a higher surface density show the formation of a vortex outside
the gap already when the planets lock into 2:1 MMR.
In the \model{PDS70 IRR} model,
the vortex survives the small migration jump and lives on until the end of the simulation.
In the \model{PDS70 ISO} model,
it lives until the migration jump when it is disrupted by the outer planet and does not form again during the simulation
time.

\begin{figure}[t]
  \centering
  \includegraphics[width=\linewidth]{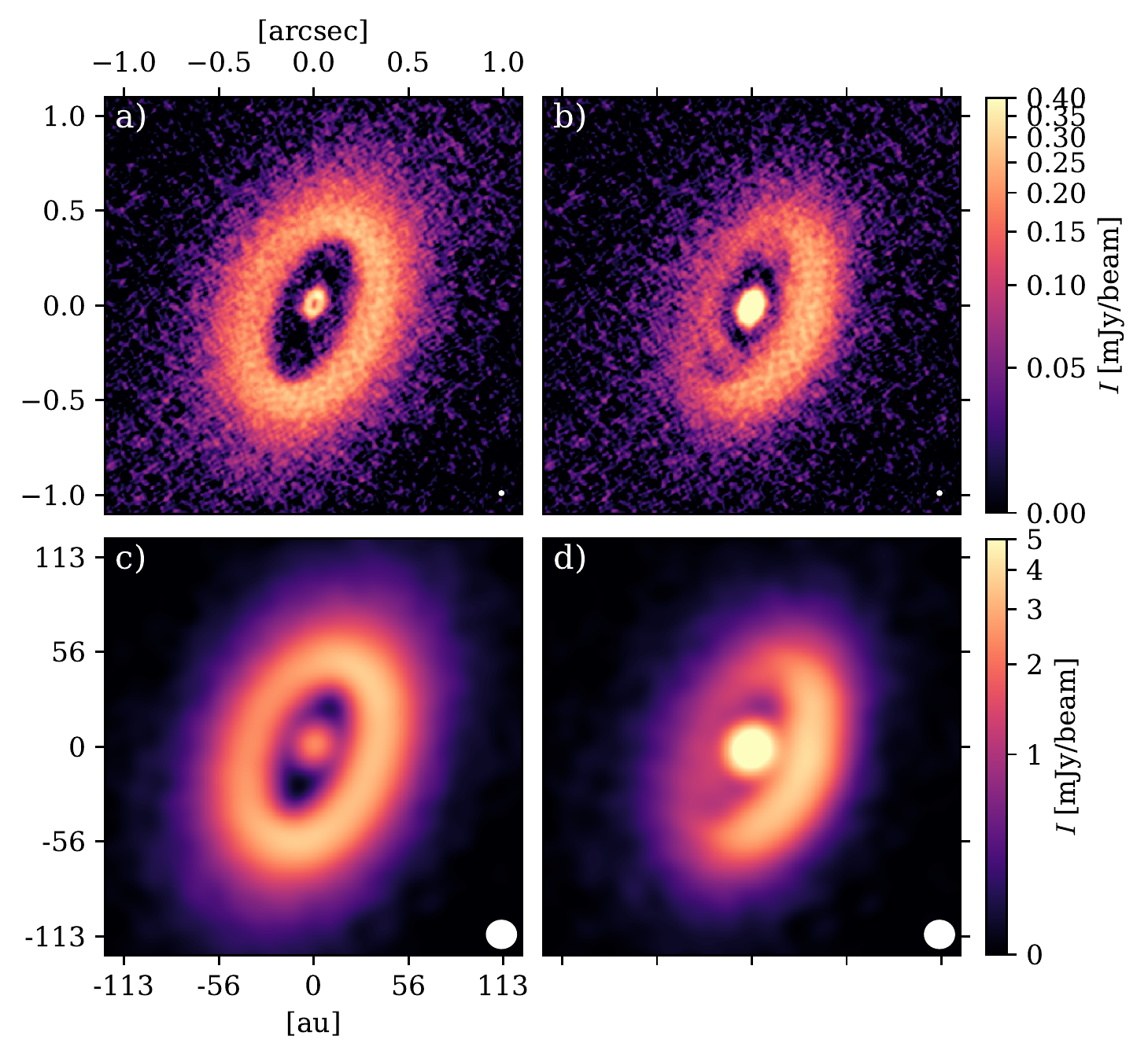}
  \caption{\label{fig:synthobs-PDS70} Synthetic observations of the PDS\,70 models.
    Disk mass increases from left to right and resolution decreases from top to bottom.
    For a lower disk mass, the appearance is smooth and symmetric, showing a large gap.
    At higher disk mass, azimuthal asymmetries appear and additional substructures emerge for higher resolution.
    The left column (a and c) shows the low disk mass model (\model{PDS70 IRR M/5}) at $t=70.3\,\text{kyr}$
    (see vertical line in Fig.~\ref{fig:migration-PDS70})
    and the right column (b and d) shows the high disk mass model (\model{PDS70 IRR}) at the same point in time.
    The top row (a and b) was generated with the same angular resolution as Fig.~\ref{fig:synth_obs_montage}
    while only the smaller ALMA antenna configuration was used for the bottom row (c and d),
    resulting in a larger beam size.
    The beam size is indicated in the bottom right corner of each image.}
\end{figure}

\subsection{Synthetic images}
Similarly, as presented in sections \ref{sec:radtrans} and \ref{sec:synthobs}, the outcomes of PDS 70 models are
post-processed in order to allow a comparison to the observed system. A modification of the procedure was
made regarding the temperature and radius of the central star which has been set to 3972 K and 1.26 $R_\odot$,
respectively \citep{2018A&A...617A..44K}.
The mass of the inner disk was reduced by a factor of $10^{-2}$ instead of $10^{-5}$ to keep some emission from the inner disk
which can be seen in the observation \citep{2019A&A...625A.118K}.
We additionally apply an inclination angle of 51.7 degrees and a position angle of 156.7 degrees.

Figure~\ref{fig:synthobs-PDS70} shows the synthetic observations of the model \model{PDS70 IRR} (panels b and d)
and the model with lower disk mass \model{PDS70 IRR M/5} (panels a and c), where the
synthetic observations are calculated using two different ALMA antenna configurations.
The first configuration (panels a and b) is unchanged to the one in Sect.\,\ref{sec:synthobs}.
The second configuration (panels c and d) only uses the smaller \texttt{alma.cycle5.5} antenna configuration which results in a larger
beam size.

Because the inner disk is more massive in gas, our assumption of a gas to dust ratio of 100 results in a very bright
inner disk for higher disk masses.
For each configuration, the colour scale is chosen such that its maximum value is the maximum intensity from the low disk mass
models (panels a and c). This is done to make the outer parts of the disk visible in both models on the same colour scale.

The low mass disk at high resolution (panel a) shows a hole in the inner disk.
This is due to the inner hole of the computational domain.
In the high disk mass case (panel b) this is not visible any more due to the selection of the maximum value for the colour scale.
For the smaller ALMA configuration with larger beam size (panels c and d), the hole is simply smeared out.

There is a clear difference between the model \model{PDS70 IRR M/5} with lower disk mass and smooth outward migration (panels a and c) and
model \model{PDS70 IRR} with higher disk mass and a migration jump (panels b and d).
The low disk mass model (panels a and c) qualitatively reproduces the dust continuum observations presented in \citet{2019A&A...625A.118K}
showing the dust ring at a large distance from the star and a clearly visible and wide gap.
Here, the location of the dust ring is closer in compared to the observations.

For a higher disk mass, the ring becomes azimuthally asymmetric with the right side being pronounced due to
existence of a vortex.
At lower resolution (panel d), only the brighter right side is visible.
At higher resolution (panel b), additional substructure is visible in the disk,
e.g. the spur feature is visible to the right of the center.
Also clearly visible is an arc like feature in the gap in the bottom left quadrant.
The arc is located closer to the center of the gap and emerges because the spiral arm of the inner planet is enhanced
in density at that point in time.
During the previous orbits, the outer planet came close to the outer gap edge on its eccentric orbit which results in
a higher mass flow across its orbit.
This higher mass flow across the outer planet's orbit subsequently leads to higher densities in the inner planet's
spiral arm.
Thus, the arc is a result of the dynamic nature of the system with its high eccentricities.

\section{Discussion}
\label{sec:discussion}

In this section, we will discuss the implications of our findings, put them into context and discuss limitations.
Results will be discussed in reversed chronological order,
starting with our PDS\,70 models (Sect.~\ref{sec:discuss_PDS70}),
followed by the synthetic observations (Sect.~\ref{sec:discuss_synthobs}),
their implications (Sect.~\ref{sec:obersational_signposts})
and migration jumps (Sect.~\ref{sec:discuss_migration_jumps}).
We then go on to discuss simulation aspects such as the role of the inner boundary condition
(Sect.~\ref{sec:discuss_boundary}) and choices concerning the equation of state and self-gravity
(Sect.~\ref{sec:discuss_numerics}).
We conclude by discussing the implications of our findings for transition disks (Sect.~\ref{sec:discuss_TDs})
and directly imaged systems of planets at large distances from their hosts stars
(Sect.~\ref{sec:discuss_distant_planets}).

\subsection{PDS 70}
\label{sec:discuss_PDS70}

Our synthetic observations of the PDS\,70 system showed a dust ring at slightly smaller radii
compared to \citet{2019A&A...625A.118K}.
If we had simulated the system for a longer time, the pair of planets would have migrated further out also pushing the location
of the ring further out.
This way the synthetic models could be fine-tuned to match the actual observations.

In the case of PDS\,70, our results suggest that no migration jump is happening in the system at this moment in time.
This is likely due to the disk mass being lower than needed for migration jumps to happen, as the comparison between
models \model{PDS70 IRR} and \model{PDS70 IRR M/5} shows.
Following this line of thought, our results might be used to put an upper limit on the disk mass.
The disk mass must be lower than the one in model \model{PDS70 IRR}, $M_\text{disk} = 0.048 \text{M}_\odot$, at the time
when the migration jump happens.
Note, that for simulations the disk mass depends on the extent of the domain and the number quoted here refer to a disk 
with the inner and outer radius of our models, $r_\text{in} = 2.08\,\mathrm{au}$ and $r_\text{out} = 208\,\mathrm{au}$.
Scaled with a power low profile as in Eq.~\eqref{eqn:ic_sigma}, this corresponds to a surface density 
of $\Sigma = 8.24\,\mathrm{g}/\mathrm{cm}^2$ at 40\,au
which is in agreement with the $\Sigma \approx 1 \,\text{g}/\text{cm}^2$ at 40\,au which was used in models of PDS70
before \citep{2018A&A...617A..44K, 2019A&A...625A.118K}.
Another upper bound could be obtained by verifying the non-existence of features due to gravitational instability
such as fragmentation which would require even more massive disks.
Our threshold is lower than this self-gravity induced upper bound because our disks have a $\Sigma$ which is lower
than the value required for fragmentation.
The Toomre $Q$ parameter is larger than 1 at any location and time.
The upper bound could likely be improved by performing a parameter study which is out of the reach of this work.

Given PDS\,70's age of $5.4\,\pm1.0\,\text{Myr}$, it can not be ruled out that a migration jump happened at earlier times,
and that the system relaxed back into a quieter state as the disk dispersed due to effects like photoevaporation or
magnetically driven disk winds \citep{2020A&A...633A..21R}.
Long term simulations including disk dispersal effects would be required to answer this question.

In our synthetic images, we also observe the spur at the outer gap edge reported in
\citet{2018A&A...617A..44K,2019A&A...625A.118K} which has already been reproduced using a similar model like ours \citep{2019ApJ...884L..41B}.
Such a feature also appeared in our generic models (see Fig~.\ref{fig:synth_obs_montage} panels b,e, and f).
This suggests that this kind of feature can generally appear for an outward migrating pair of planets.

\subsection{Synthetic observations and dust treatment}
\label{sec:discuss_synthobs}

In Sect.~\ref{sec:synthobs}, we modelled dust emission by assuming a uniform dust to gas mass ratio of $10^{-2}$.
Thus, we use gas dynamics as a proxy for dust dynamics.
As a result, effects like dust drift, dust size filtration and dust diffusion
\citep{2018ApJ...854..153W} which probably are at play in a protoplanetary disk are not considered.
The inclusion of a proper dust treatment might therefore change the appearance of some features in the synthetic images.

Nonetheless, similar features have already been observed to also emerge when dust is handled properly.
A dust ring growing in size due to outward migration of a pair of massive planets was reported in
\cite{2019AJ....157...45M} for a close-in (inside 10\,au) Jupiter-Saturn pair in a massive disk
($\Sigma_0(10\,\text{au}) = 75\,\text{g}/\text{cm}^2$ vs. $\Sigma_0(10\,\text{au}) \approx 46\,\text{g}/\text{cm}^2$ here)
and by \citet{2019ApJ...884L..41B} for a model of the PDS\,70 system with $10$ and $2.5\,M_\text{Jup}$ planets in a lighter disk
($\Sigma_0(10\,\text{au}) \approx 9\,\text{g}/\text{cm}^2$).
The latter study also found the PDS\,70 spur feature \citep{2019A&A...625A.118K,2019ApJ...884L..41B},
supporting that similar features also appear with proper dust treatment.
However, to judge how dust dynamics influences the features found in our synthetic observations,
simulations including dust treatment are needed.
We plan to investigate this in a follow-up study.

We want to highlight, that multiple distinct observational features can be created by the same planetary system
at different points in time, if the system exhibits strong enough dynamic effects such as a migration jump.
Except for the choice of planet mass ratio,
all presented observational features emerged for a disk with standard parameters without the need of fine-tuning.

\subsection{Observational signposts for dynamic effects}
\label{sec:obersational_signposts}

For our PDS\,70 models (see Sect.\ref{sec:PDS70}),
we observed non-axisymmetric features in the higher disk mass model.
The higher surface density caused faster outward migration and thus a higher mass flow through the common gap.
The first effect was the appearance of a vortex, visible as asymmetry in the intensity distribution
(Fig.~\ref{fig:synthobs-PDS70} panel b and d).
The second effect was an arc like feature in the gap region
caused by an enhancement of density in the spiral arm of the inner planet.
Together with the already discussed spur feature, if detected in real observations,
these feature might hint at outward migration and enhanced mass flow through the gap region.

Contrasting the smooth synthetic observations of our low mass PDS\,70 model or the one from \citet{2019ApJ...884L..41B},
in which the planetary system undergoes smooth and slow outward migration, with our synthetic observations in
Fig.~\ref{fig:synth_obs_montage}, we can identify which features could be signposts for strong dynamical effects like
migration jumps.
We can identify these potential signposts as very eccentric holes or gaps, vortices,
arc shaped voids, mass accumulations in Lagrangian points and visible spiral arms.
The latter two might be the easiest to discover in real observations because they are located in the gap region where
the surrounding emission is weak.

Indeed, similar features such as an arc inside a gap have already been observed, e.g. in the HD~163296 disk
\citep{2018ApJ...869L..49I}.

Fast outward migration of a single planet might lead to comparable observational signatures 
such as arc shaped voids behind the planet \citep{2008MNRAS.387.1063P}.
Therefore, only the combination of one of signposts listed above with a large gap will provide a strong indication
for a migration jump.

Our simulations show that a higher disk mass facilitates stronger planetary system dynamics.
Thus, by pointing toward strong dynamic effects, these signposts might indirectly hint at a high gas mass of the disk.

\subsection{Migration Jumps}
\label{sec:discuss_migration_jumps}

Migration jumps were introduced in Sect.~\ref{sec:migration_jumps}.
To our knowledge, we are first to name, describe and analyze this dynamical process in detail.
We observed migration jumps for different resolutions (Appendix~\ref{sec:resolution_test}),
for different choices of equation of state and treatment of the energy equation
(Sect.~\ref{sec:PDS70} and Appendix~\ref{sec:depence_eos}),
for different domain sizes (models \model{L} and \model{L M/2}),
independent of accretion onto the planet (Sect.~\ref{sec:accretion})
and for different choices of inner boundary conditions (models \model{VB*} and \model{WD*})
and for models with an without self-gravity (models \model{SG} and \model{SG IRR}).
These tests make us very confident, that migration jumps are indeed a physical effect.

There is at least one example in the literature where migration jumps appeared in simulations.
Fig.~1 of \citet{2020MNRAS.492.6007C} shows two migration jumps in their model \model{7b} in which a pair of Jupiter
and Saturn (mass ratio 3:1) is migrating outwards in 3:2 MMR and experiences two small migration jumps
between 5 and 8\,au.
This indicates, that migration jumps can also happen for other types of mean motion resonances, at smaller radii and
for planet masses down to a pair of Jupiter and Saturn.
A natural question to ask is: why migration jumps were not found before, in studies like
\citet{2019AJ....157...45M} or \citet{2019ApJ...884L..41B}?

For the case of \citet{2019AJ....157...45M}, we repeated similar simulations with different
boundary conditions and the domain size from 0.5\,au to 15\,au.
We found smooth outward migration at similar rates.
In these simulations, the surface density, $\Sigma$, is about twice as high as in our model \model{M9-3}.
The non-appearance of a migration jump is likely due to the non-existence of a vortex in these simulations.
This is probably due to the smaller outer radius and the wave damping zone close to the outer boundary which prevents
the formation of a vortex at the location where we would expect it by scaling down the location from our simulations.
This hints at the importance of vortices for the mechanism of migration jumps.

In the case of \citet{2019ApJ...884L..41B}, the non-appearance of migration jumps is very likely due to the lower value
$\Sigma$, which is comparable to our model \model{M9-3 M/10} which also does not show a migration jump.

Migration jumps could have profound effects for planetary systems.
During the jump, dust can be gravitationally scattered by the planets which might be a way to redistribute dust
trapped in a pressure maximum and possibly even dust trapped in the vortex.
We plan to perform simulations with embedded dust to check these hypotheses.

A jump might also have significant effects for the accretion process as shown in Sect.\ref{sec:accretion}.
Mass accretion rates onto the planets can be enhanced by two orders of magnitude by moving the outer planet to regions
outside the planet gap where the surface density is high.
This in turn also increases the mass accretion onto the inner planet.
Together this might provide a mechanism how massive planets can accrete mass more efficiently by tapping into mass
reservoirs far away from their initial orbit.

In our view, migration jumps are a composite phenomenon in which
resonant outward migration via the \cite{2001MNRAS.320L..55M} mechanism,
the interaction of a planet with a vortex, 
and the subsequently triggered type III rapid outward migration 
are combined to give rise to an emerging effect.

\subsection{Planet ejections and internal boundary condition}
\label{sec:discuss_boundary}

Planet ejections occurred in models where an inner boundary different from outflow was used in combination with a
disk centered in the center of mass of the N-body system due to a numerical instability (Sect.~\ref{sec:planet_ejection}).

Our results suggest that if a viscous boundary, a reflective boundary, or wave damping zones are used,
the hydrodynamical simulation should be centered onto the primary star.
It might also be possible to adjust the boundary condition to follow the moving star in the other case, but we
did not implement this more complicated feature.

Such a dependence of the dynamical behaviour of planets on the treatment of the inner boundary,
which lies well inside the actual realm of the planets,
has been observed in other simulations of embedded planets, for example for the system GJ\,876
\citep{2008A&A...483..325C,2018A&A...618A.169C}.

The outflow boundary condition seems to give more freedom to the inner disk by allowing a moving inner disk edge.
Indeed, this boundary condition is also a good choice to simulate eccentric disks around binary stars
\citep{2018A&A...616A..47T}.

\subsection{Equation of state and self-gravity}
\label{sec:discuss_numerics}

Recent studies showed, that radiative effects can play an important role for the spiral arm and gap structure in the case of
low mass planets, where radiative effects cause significant change compared to a purely locally isothermal assumption
\citep{2020A&A...633A..29Z}.
In our case we saw a qualitatively similar behaviour for locally isothermal simulations and simulations considering
radiative effects (see Appendix~\ref{sec:depence_eos}).
However, the size of migration jumps and the rate of outward migration depended on the inclusion of
radiative effects in our PDS\,70 models (Sect.~\ref{fig:migration-PDS70}).
The appearance of migration jumps for both, radiative and locally isothermal, models hints at the dynamic nature of the
process.
Migration jumps are likely dominated by resonant N-body interaction pumping eccentricities and the interaction with
the vortex formed outside the gap.

For reasons of simplicity and runtime, we neglected self-gravity in most of our simulations.
Judging by the value of the Toomre $Q$ parameter which stays above 1 at all times, the disk is not prone to fragmentation.
However, other processes might be altered slightly when self-gravity is taken into account.
Firstly, self-gravity might play a role for the migration of massive planets as it is the case for low mass planets
\citep{2020A&A...635A.204A}.
Secondly, in our simulations, the occurrence of migration jumps coincides with the existence of a vortex.
When self-gravity is considered, vortices will weaken and can stretch out even for low mass disks as long as the 
Toomre $Q$ parameter is lower than 50 or $h\,Q \lesssim \frac{\pi}{2}$ 
\citep{2013MNRAS.429..529L,2017MNRAS.471.2204R, 2016MNRAS.458.3918Z}.
Both conditions are fulfilled in the standard \model{M9-3} model with $Q \approx 6$ and $h\,Q \approx 0.3$ for the
initial profile at the location of the vortex.

Models \model{SG} and \model{SG IRR} with self-gravity enabled indeed show that there is no disk fragmentation,
but stretching of the vortex.
Hence, our finding that migration jumps still occur with self-gravity considered, 
is one more indication that they are indeed a physical phenomenon.

\subsection{Mass accretion and Type II transition disks}
\label{sec:discuss_TDs}

In all our models, mass accretion onto the star, as measured by the mass flow rate through the inner boundary, was
higher ($10^{-8}$ to $10^{-7} \, \text{M}_\odot/\text{yr}$) than the viscous mass accretion rate of the unperturbed disk
($<10^{-8} \, \text{M}_\odot/\text{yr}$).
The inner disk was building up mass over time in these models,
showing higher surface densities compared to the initial profile.
We suspect that mass transfer through the gap is enhanced by the 'shoveling' mechanism (Sect.~\ref{sec:accretion}).
As the outer planet comes close to the outer gap edge at apastron on its eccentric orbit,
it scatters mass inwards.
Additionally, outward migration must necessarily enhance mass flow through the gap.
When the planets migrate outward, they gain angular momentum.
Because angular momentum is conserved, the angular momentum gained by the planets has to be extracted from the gas
requiring that some gas has to move from outside the gap (outside the planets' orbits) to inside the gap
(inside the planets' orbits) to supply the angular momentum.
This mechanism of shovelling matter from outside in is after all the basic mechanism behind the Masset-Snellgrove mechanism
of outward migration, as the material crossing the joint gap is collected in the inner disk which generates the positive torque to drive
the planets outward \citep{2001MNRAS.320L..55M}. Our models serve to quantify this process in more detail.

A particularly clear example of this mechanism is model \model{L} for which we reported the dependency of mass accretion
through the gap on the direction of migration in Sect.~\ref{sec:accretion}.
There, mass flow through the gap was present during both in- and outward migration but was strongly enhanced for the outward case.
This means, that the common gap formed by the planets is not an impermeable barrier for mass accretion.
Model \model{L} illustrates that mass flow through the gap is possible for both direction of migration and that it can
be enhanced by over one order of magnitude in case of outward migration.

The large gap in dust emission reported in Sects.~\ref{sec:observability} and \ref{sec:PDS70}
\citep[see also][]{2019AJ....157...45M,2019ApJ...884L..41B}
together with an enhanced stellar mass accretion make outward migrating pairs of planets a prime candidate for a
consistent explanation of Type II transition disks \citep{2012MNRAS.426L..96O} which feature large gaps or holes and
high accretion rates at the same time.

\subsection{Directly imaged planets at large distances}
\label{sec:discuss_distant_planets}

There are several examples of giant planets observed at large distances to their host star.
At such large distances, in situ formation by gravitational instability is challenging \citep{2012ApJ...746..110Z}.
Assuming they formed further in,
it is still challenging to explaining how they moved outwards over long distances, even though models based
on the smooth outward resonant migration have been invoked to explain such systems
\citep{2008MNRAS.387.1063P,2009ApJ...705L.148C,2020A&A...633A...4K}. The advantage of our migration jumps is the very short
timescale and the large radial range covered.

PDS\,70 is one example with the outer planet located at $35.5\pm2\,\text{au}$ \citep{2019NatAs...3..749H} for which
outward migration seems a promising scenario as a comparison with our simulations shows.
In our simulations the outer planet reached distances from its host star of up to 133\,au after 226\,kyr, providing an
explanation of how planets can reach such large distances from their host after being formed further inside.
The actual process is likely a balance between speed of migration and dispersal of the disk, allowing for a range of
final locations.

HR~8799 is another famous example of a planetary system directly observed which features a chain of four planets which might be
in 8:4:2:1 resonance \citep{2010Natur.468.1080M,2014MNRAS.440.3140G}. Here, with the outer planet HR~8799~b is
located around 70\,au \citep{2018AJ....156..192W}.
It is unclear, whether outward migration in resonance can also produce such an intricate system, yet, formation
of the planets closer to the star followed by outward migration should be considered as a formation scenario.
To answer this question, more extended simulations with more planets will have to be performed in the future.

\section{Summary}
\label{sec:summary}
We studied the dynamical evolution of a system of two massive planets (in a mass range of 3 to 9 $M_\mathrm{Jup}$)
embedded in a protoplanetary disk using two-dimensional,
viscous hydrodynamical simulations carried out with the FARGO-code.
The planets were treated as smoothed point masses that in some simulations were allowed to accrete mass
which was added to their dynamical mass.
For the disk we assumed either a locally isothermal equation or a more realistic situation
where we solved for an energy equation that included viscous heating,
radiative cooling and stellar irradiation.

Concerning the migration of the planets we found two different basic behaviours,
depending on the mass order of the two planets.
In the case of a more massive outer planet,
the planets were migrating inward engaged in 2:1 mean-motion resonance (MMR).
For a more massive inner planet with mass ratios of 2:1 or 3:1, we found outward migration of both planets,
again engaged in a 2:1 MMR.
As found before,
this resonant migration process,
originally described by \citet{2001MNRAS.320L..55M} for the Jupiter-Saturn system,
can lead to a resonant outward migration in the 2:1 MMR in the case of massive planets
\citep{2008MNRAS.387.1063P,2009ApJ...705L.148C}.

The new feature that we discovered is an occurrence of phases of rapid outward {\it migration jump}
of the outer (lighter) planet where the planet covers a large radial distance in a very short timescale,
for example from 40 to 72\,au in only 5\,000 years, see Fig.~\ref{fig:migration-sample-zoomin}.
They are a composite phenomenon in which outward migration in resonance and interaction with a vortex
cause conditions such that type III rapid outward migration can be triggered.
The migration jumps are a generic, robust feature of our models.
They occur for different equations of state and accretion rates onto the planet, 
with or without self-gravity, and for different resolutions,
as long as the surface density of the disk is high enough.

In addition to the dynamical behaviour of the embedded planets we monitored the mass accretion onto the central star,
as this is a standard observable feature in transition disks \citep{2016PASA...33....5O}.
For our models we find that during the smooth outward migration phase of
the resonantly locked planet pair the accretion rate is significantly higher than in the situation when the outer planet has a higher mass
and the planets migrate inward. During the outward migration phases the planetary system gains angular momentum which is lost by the
disk.
As a consequence the disk material moves inwards and is 'shovelled' towards the star by the pair of planets.
This increases the mass flow rate onto the star by more than one order of magnitude,
much higher than in the regular phases of smooth inward migration.
It can be even more enhanced during the short phase of a migration jump,
see Figs.~\ref{fig:accretion-accreting} and \ref{fig:orbit-swap}.
A combination of outward migration and high stellar mass accretion,
as found in our models, could serve as a consistent explanation for the phenomenon of Type II transition disks 
with large inner holes and nevertheless high stellar accretion rates \citep{2012MNRAS.426L..96O}.

Using the outcome of our hydrodynamical models we calculated synthetic images that show a surprising variety of
non-axisymmetric features appearing over time in a single system, see Fig.~\ref{fig:synth_obs_montage}.
Depending on their dynamical state,
a bright ring just beyond the planets was seen,
then vortex structures,
and then additional structures in the main gap created by the planets.
These initial images were based on a constant dust to gas ratio
and for more realistic cases the dust dynamics will have to be followed simultaneously to the gas dynamics.
Nevertheless, the initial models give an insight of the possible observational effects generated by the planets.
In our study,
we also included models with parameters geared to the system PDS\,70, which contains two massive embedded planets.
From our synthetic images for this system we may conclude that it does not currently undergo a migration jump but
might very well be in a phase of outward migration, compatible with \citet{2019ApJ...884L..41B}.
The non-occurrence of the signposts of a migration jump indicates a disk mass lower than 
$M_\text{disk} < 0.048\,\mathrm{M}_\odot$ for a disk extending out to 200\,au which serves an independent upper bound compatible with
radiative transfer models reproducing the observations of the system \citep{2018A&A...617A..44K,2019A&A...625A.118K}.
A more detailed comparison between simulations and observations will have to consider dust embedded in the disk.

From our models we can finally conclude that Type II transition disks with large inner holes but significant stellar accretion
are indeed signpost for highly dynamic embedded planetary systems.

\begin{acknowledgements}
  All authors acknowledge funding from the DFG research group FOR 2634 ''Planet Formation Witnesses and Probes: Transition Disks''
  under grant DU 414/22-1 and KL 650/29-1, 650/30-1.
  The authors acknowledge support by the High Performance and Cloud Computing Group at the Zentrum
  f\"ur Datenverarbeitung of the University of T\"ubingen, the state of Baden-W\"urttemberg
  through bwHPC and the German Research Foundation (DFG) through grant INST\,37/935-1\,FUGG.
  Plots in this paper were made with the Python library \texttt{matplotlib} \citep{hunter-2007}.
\end{acknowledgements}

\bibliographystyle{aa}
\bibliography{migration_jumps}

\appendix

\section{Convergence with resolution\label{sec:resolution_test}}

To test convergence of the simulations with respect to spatial resolution,
we ran a model with the 2D resolution lowered and increased by a factor of 2, i.e.
in each direction the resolution is changed by a factor of $\sqrt{2}$.
Model \model{M9-3 HR} has $N_r \times N_\phi = (426, 580)$ and model \model{M9-3 DR} has
$N_r \times N_\phi = 851 \times 1161$ cells.
They are otherwise identical to the \model{M9-3} model which has $N_r \times N_\phi = 602 \times 821$ cells.

\begin{figure}[t]
  \centering
  \includegraphics[width=\linewidth]{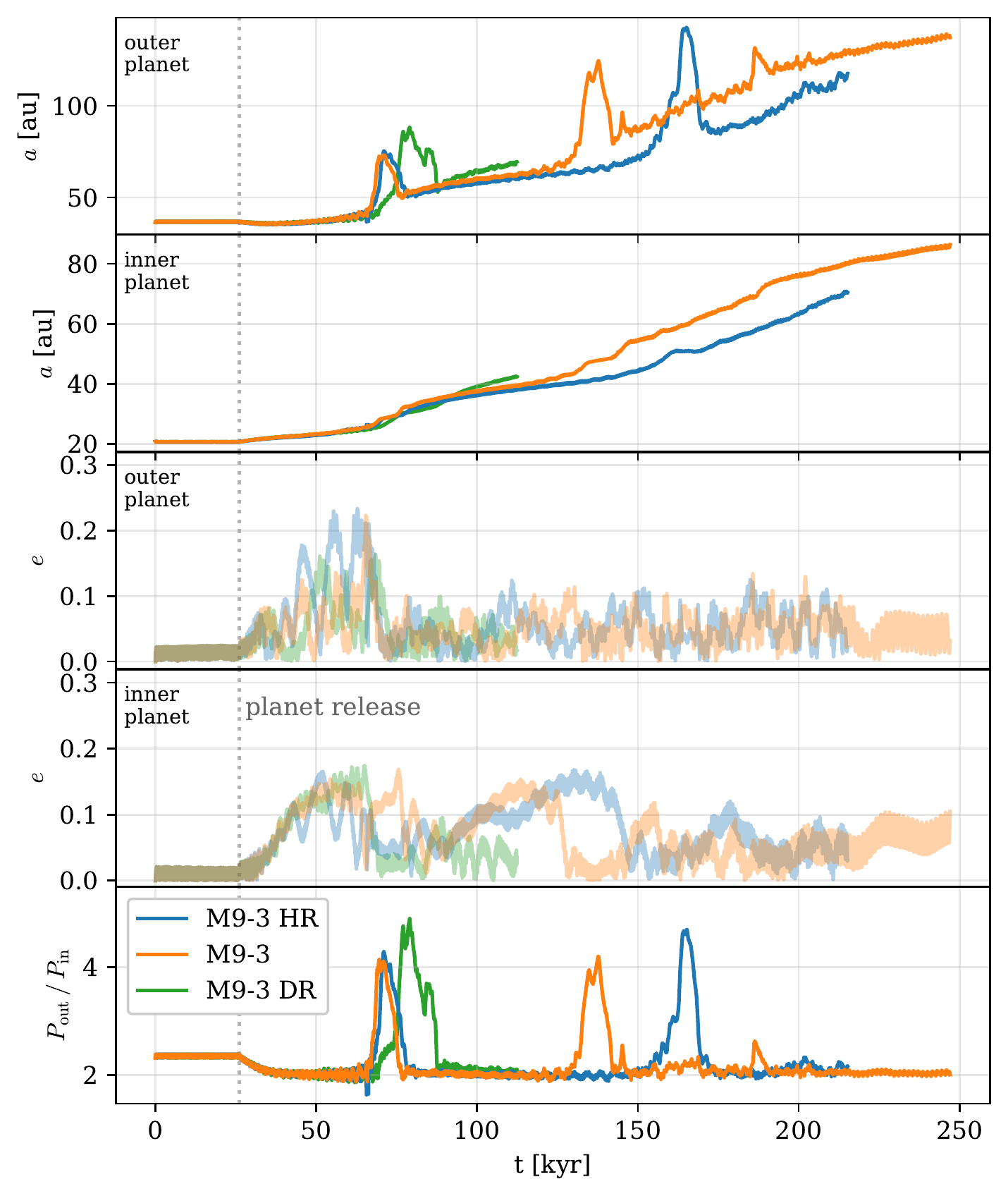}
  \caption{\label{fig:migration-resolution} Migration history of models with standard, double (\model{DR})
    and half (\model{HR}) 2D resolution to test convergence with spacial resolution.
    The panels show, from top to bottom, the evolution of: semi-major axis of outer and inner planet (top and \second),
    their eccentricities (\third and \fourth) and their period ratio (bottom).}
\end{figure}

A comparison of the planet migration for the three cases is shown in Fig. \ref{fig:migration-resolution}.
The overall qualitative behaviour is the same, the most noticeable difference is the time at which the migration jump occurs.
For the first jump this is only slightly different.
Since the changes are small, we conclude that the simulations are converged with respect to resolution.
In fact, the lower resolution is already sufficient to resolve the dynamics.

\section{Dependence on equation of state}
\label{sec:depence_eos}

\begin{figure}[t]
  \centering
  \includegraphics[width=\linewidth]{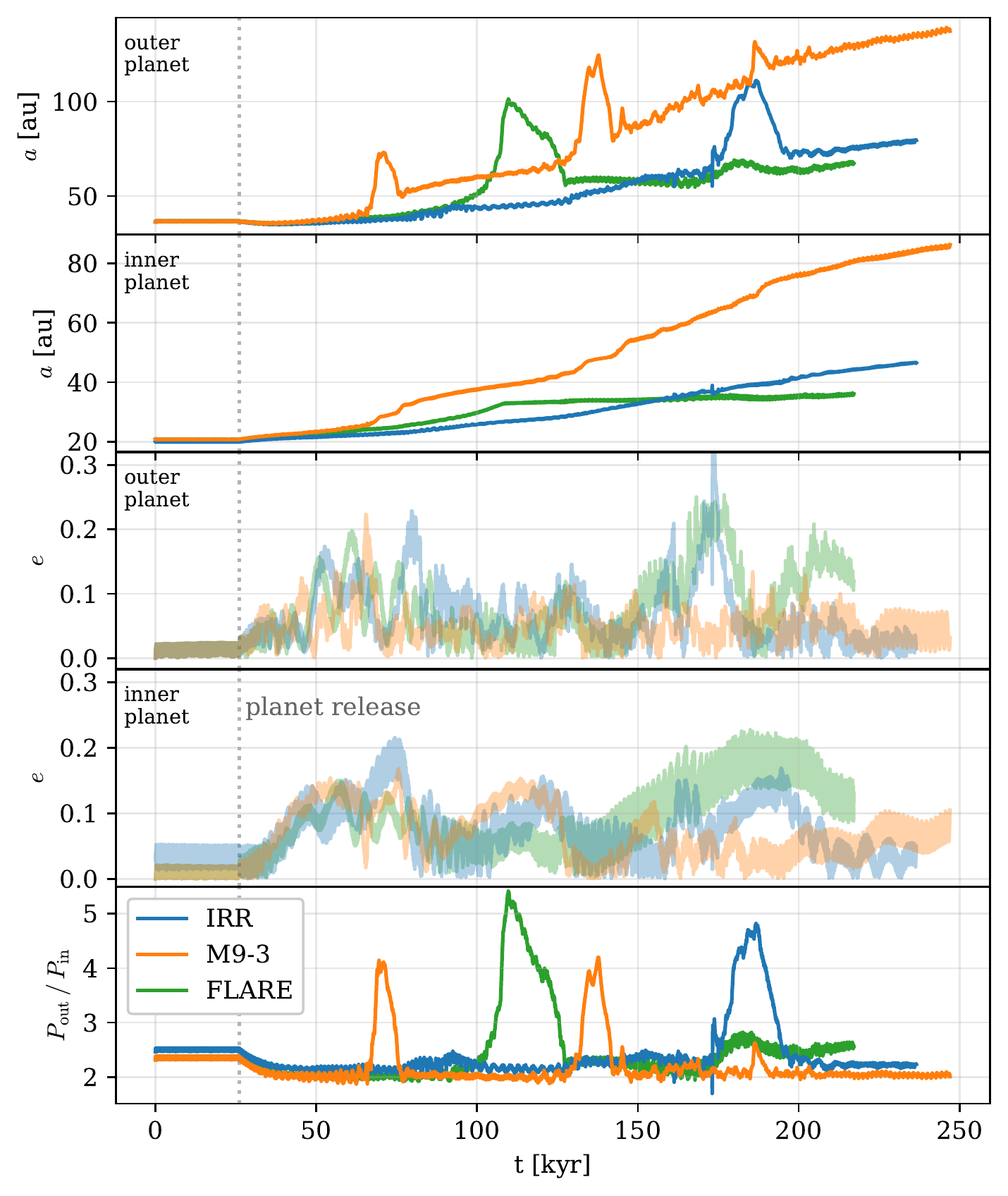}
  \caption{\label{fig:migration-irradiation}Comparison of the migration history of the standard
    \model{M9-3} model and the more realistic and radiative model \model{IRR} and the model \model{FLARE}.
    Panels as in Fig.~\ref{fig:migration-resolution}.}
\end{figure}

Using the locally isothermal equation of state is a valid approximation in the case of low optical thickness and
negligible viscous heating compared to stellar irradiation.
For the outer disk region, in which the planets in our simulations are located, this should be well justified.
To test the assumption, we ran additional simulations where the energy equation is solved and viscous heating and
cooling from the disk surfaces are included as in \cite{2012A&A...539A..18M}, and irradiation from the star is
treated analogous to \cite{2020A&A...633A..29Z} with their choices of parameters.
Additionally, we repeated the standard \model{M9-3} with a radial temperature profile that is set by stellar irradiation,
which resembles the temperature profile in model \model{IRR} very well resulting in a flared disk
(see Fig.~\ref{fig:hoverr-irradiation} below).

Figure~\ref{fig:migration-irradiation} shows a comparison of the migration history
between the reference model \model{M9-3} and the irradiation model \model{IRR}.
Although the outward migration happens with a slower speed, a migration jump still occurs.
Model \model{IRR} shows a first event at around $80\,\mathrm{kyr}$ where the eccentricity of the outer planet suddenly
drops from 0.2 down to small values, but only jumps about $5\,\mathrm{au}$.
A second event occurs around $170\,\mathrm{kyr}$ where the outer planet jumps $45\,\mathrm{au}$, comparable to
the migration jumps observed in the M9-3 run.
Model \model{FLARE}, which employs a locally isothermal equation of state, first follows the \model{IRR} model but instead
of the failed small jump at $80\,\mathrm{kyr}$ it undergoes a full large jump.
The comparison shows that, although details as the value of migration rate depend on the aspect ratio profile,
migration jumps appear also for flared disks.

Figure~\ref{fig:hoverr-irradiation} shows the temperature and aspect ratio profiles for the three models.
Around the region where the planets are located, the aspect ratios are comparable with values around 0.05.
However, at the location where the planet jumps happens, $h$ has increased to $\sim 0.07$ 
for the \model{IRR} and \model{FLARE} models.
This might explain the difference in amplitude and period ratios during the migration jump.
Finally, since there are only small qualitative differences,
we conclude that the locally isothermal assumption is justified in this case to capture the most important dynamics.

\begin{figure}[t]
  \centering
  \includegraphics[width=\linewidth]{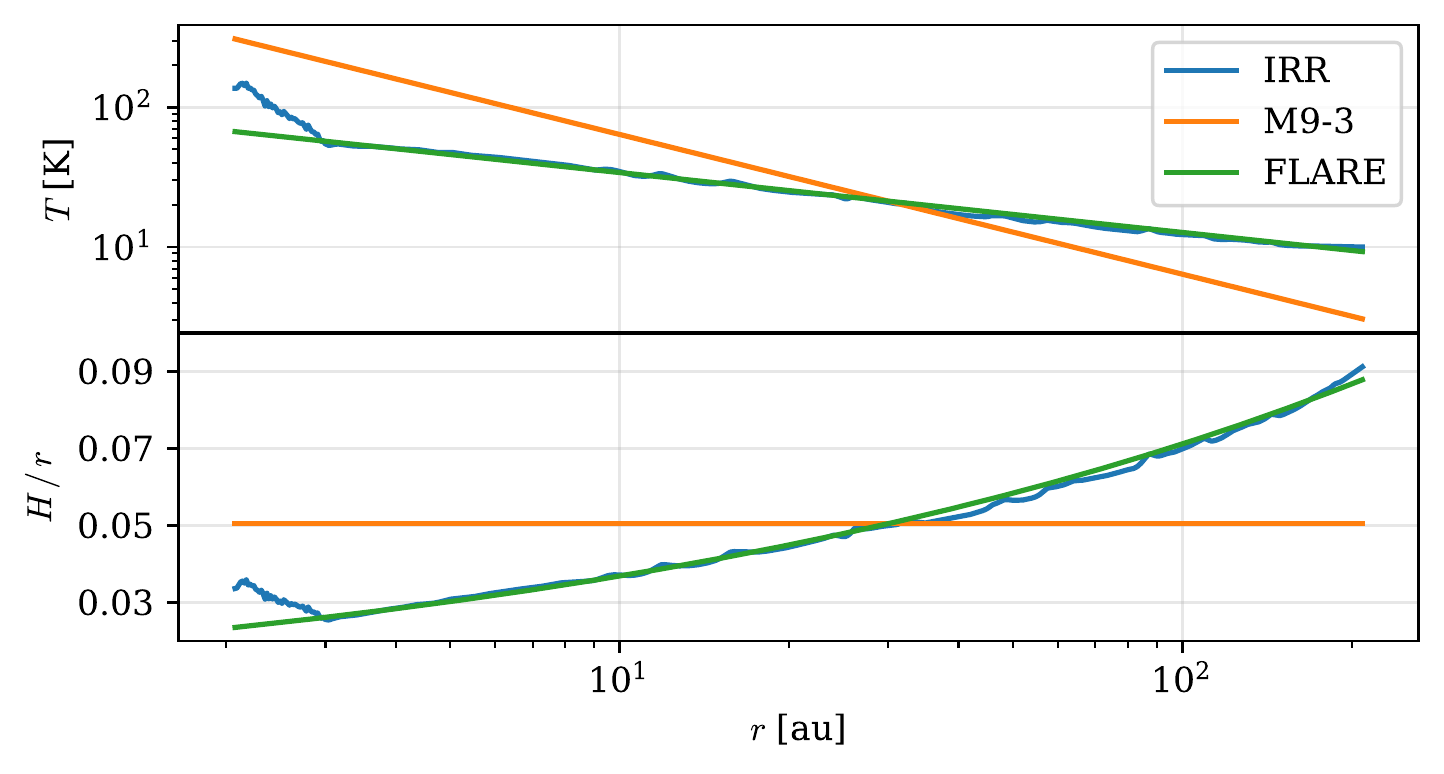}
  \caption{\label{fig:hoverr-irradiation} Temperature (top) and aspect ratio $H\,/\,r$ (bottom) 
  for the standard model \model{M9-3}, model \model{IRR} and model \model{FLARE} at 200\,kyr.}
\end{figure}

\section{Identification of the Vortex}
\label{sec:app_vortex}

During the periods of resonant outward migration, a banana-shaped overdensity appears just outside the common gap.
In this section, 
we analyse the snapshot in panel b of Fig.\ \ref{fig:2d_surface_density} of model \model{M9-3} in more detail.

\begin{figure}[t]
    \centering
    \includegraphics[width=\linewidth]{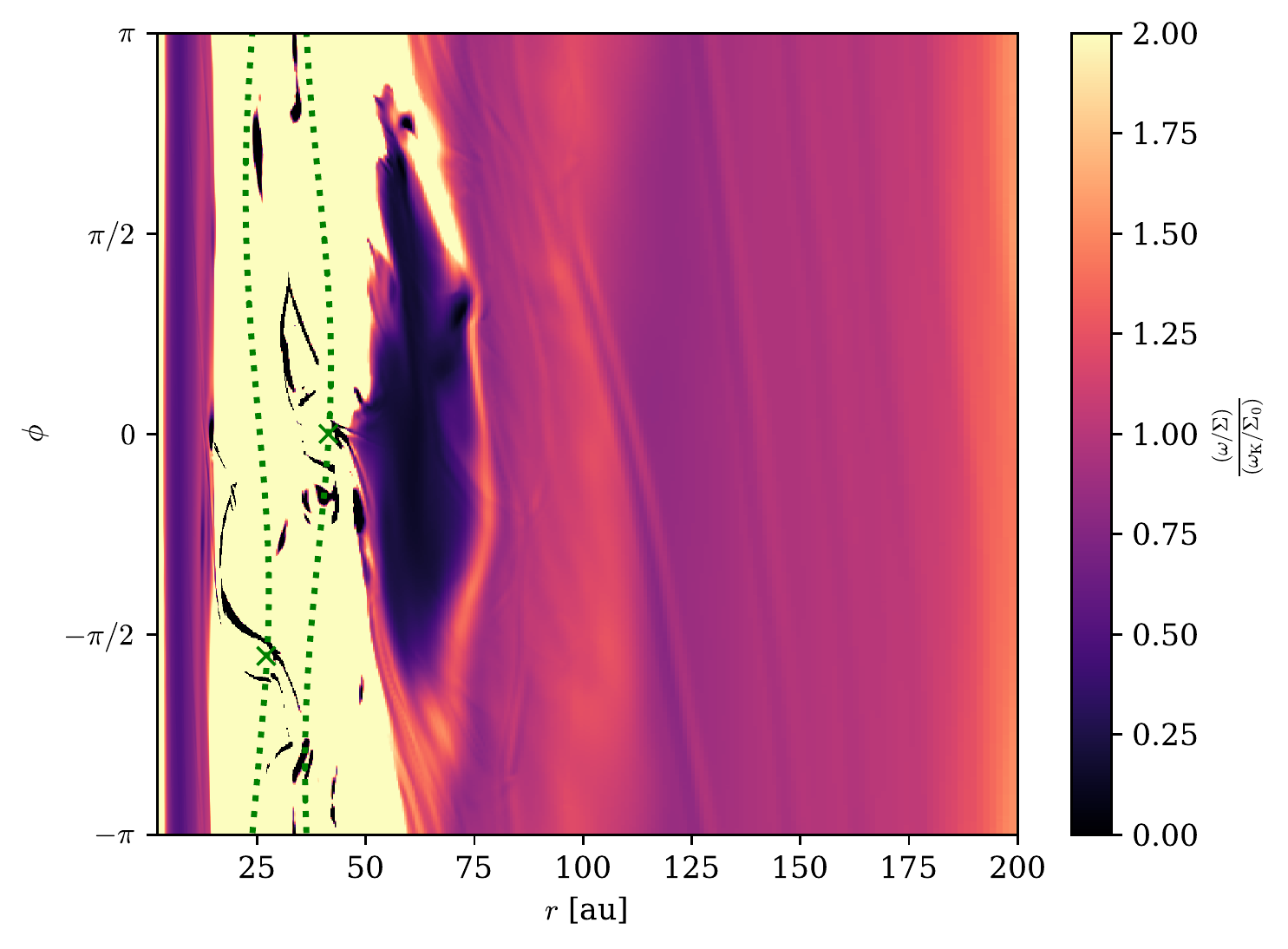}
    \caption{Vortensity normalized by the initial vortensity in $r-\phi$ coordinates for a $\pm 100\,\mathrm{au}$ zoom-in 
      to panel b of Fig.\ \ref{fig:2d_surface_density}. 
      The overdensity appears as a region of lower vortensity compared to
      its surroundings indicating that it is indeed a vortex.
      The orbits of the planets are shown as dotted green lines and each planet's location is indicated by a green $\times$.
      \label{fig:vortensity}}
 \end{figure}

Fig.\ \ref{fig:vortensity} shows the vortensity $\omega / \Sigma$ normalized by its value from the initial profile.
The numerator is the vorticity which is defined as the $z$-component of the curl of the velocity,
$\omega = (\nabla \times \vec{v})_z$. 
Both planets' orbits are shown as a dotted green line and the planets' locations are indicated by the green crosses.

By dividing the current value of the vortensity by the initial one at each location, the dependence on the steepness of the initial density 
profile and the background vorticity of the Keplerian disk are factored out.
This results in a clear picture of what happens locally with the velocity field.

A vortex in a disk appears as a region of lower vortensity as compared to its surroundings, due to its anticyclonic
nature as seen in Fig.\,\ref{fig:streamlines}.
This clearly identifies the overdensity at 59\,au as a vortex.

\FloatBarrier

\end{document}